\documentclass{aa}
\usepackage{graphicx}
\usepackage{bm}
\usepackage{txfonts}
\usepackage{natbib}
\usepackage[colorlinks,allcolors=blue]{hyperref}
\bibpunct{(}{)}{;}{a}{}{,}
\graphicspath{{./fig/}{./png/}}
\usepackage{xcolor}

\newcommand{\DD}{{\rm D}}

\newcommand{\mUUi}{\overline{U}_i}

\newcommand{\mBBi}{\overline{B}_i}

\newcommand{\mBBr}{\overline{B}_r}

\newcommand{\Eqsa}[2]{Eqs.~(\ref{#1}) and (\ref{#2})} 
\newcommand{\Equsa}[2]{Eqs.~(\ref{#1}) to (\ref{#2})}
\newcommand{\EQ}{\begin{equation}}
\newcommand{\EN}{\end{equation}}
\newcommand{\EQA}{\begin{eqnarray}}
\newcommand{\ENA}{\end{eqnarray}}
\newcommand{\brac}[1]{\langle #1 \rangle}
\newcommand{\pd}{\partial}

\newcommand{\mean}[1]{\overline{#1}}

\newcommand{\cP}{c_{\rm P}}
\newcommand{\cV}{c_{\rm V}}

\newcommand{\etat}{\eta_{\rm t}}

\newcommand{\etaT}{\eta_{\rm T}}

\newcommand{\urms}{u_{\rm rms}}

\newcommand{\ustar}{u_\star}

\newcommand{\Urms}{U_{\rm rms}}

\newcommand{\Brms}{B_{\rm rms}}

\newcommand{\chiSGS}{\chi_{\rm SGS}}

\newcommand{\Co}{{\rm Co}}

\newcommand{\CoF}{{\rm Co}_{\rm F}}

\newcommand{\Pe}{{\rm Pe}}

\newcommand{\Pra}{{\rm Pr}}

\newcommand{\PraSGS}{{\rm Pr}_{\rm SGS}}

\newcommand{\PrM}{{\rm Pr}_{\rm M}}

\newcommand{\RaF}{{\rm Ra}_{\rm F}}
\newcommand{\RaFS}{{\rm Ra}_{\rm F}^\star}

\newcommand{\Rey}{{\rm Re}}

\newcommand{\ReM}{{\rm Re}_{\rm M}}

\newcommand{\Ta}{{\rm Ta}}

\newcommand{\Mij}{M_{ij}}

\newcommand{\Mrp}{M_{r\phi}}
\newcommand{\Mtp}{M_{\theta\phi}}

\newcommand{\Rij}{R_{ij}}

\newcommand{\Rrp}{R_{r\phi}}
\newcommand{\Rtp}{R_{\theta\phi}}
\newcommand{\tij}{T_{ij}}

\newcommand{\Trp}{T_{r\phi}}
\newcommand{\Ttp}{T_{\theta\phi}}

\newcommand{\mOm}{\mean{\Omega}}

{}

\newcommand{\calK}{{\cal K}}

\newcommand{\calR}{{\cal R}}
\newcommand{\calS}{{\cal S}}

\newcommand{\rin}{r_{\rm in}}

\newcommand{\FFFrad}{{\bm F}^{\rm rad}}

\newcommand{\Fbot}{F_{\rm bot}}

\newcommand{\nad}{n_{\rm ad}}

\def\onethird{{\textstyle{1\over3}}}

\def\onehalf{{\textstyle{1\over2}}}

\def\threehalfs{{\textstyle{3\over2}}}

\newcommand{\Figa}[1]{Fig.~\ref{#1}}

\newcommand{\Fig}[1]{Figure~\ref{#1}} 

\newcommand{\Sec}[1]{Section~\ref{#1}} 

\newcommand{\Table}[1]{Table~\ref{#1}}

\definecolor{ForestGreen}{RGB}{34,139,34}
\definecolor{AGray}{rgb}{.4,.4,.4}
\definecolor{LightYellow}{rgb}{1.,1.,.8}
\definecolor{LightCyan}{rgb}{0.88,1,1}
\begin{document}

\authorrunning{K\"apyl\"a}
\titlerunning{Grand minima and intermittency in near-critical convection-driven dynamos}

   \title{Grand minima and intermittency in near-critical\\ convection-driven dynamos}

   \author{P. J. K\"apyl\"a
          \inst{1,2}
          }

   \institute{Institute of Physics, University of Graz, 8010 Graz, Austria 
              email:
              \href{mailto:petri.kaepylae@uni-graz.at}{petri.kaepylae@uni-graz.at}
   \and Institut f\"ur Sonnenphysik (KIS), Georges-K\"ohler-Alle 401a, 79110 Freiburg
              im Breisgau, Germany
            }

\date{\today}

\abstract{Solar magnetism exhibits long-term variations on timescales
  of decades to centuries.}%
{The main aim is to study how the supercriticality of the dynamo
  affects the variability of magnetic fields on timescales relevant to
  the Sun.}%
{Three-dimensional rotating magnetohydrodynamic simulations in a
  spherical wedge geometry are used to produce self-consistent
  convection-driven dynamos. The supercriticality of the dynamo is
  controlled by varying the magnetic diffusivity.}%
{The main result is that long-term variability of the dynamo increases
  with decreasing supercriticality of the dynamo and that cases
  nearest to the onset of dynamo action show extended grand
  minima-like periods of low activity. The simulations fall in three
  classes where the least supercritical cases show the largest
  variations of magnetic activity and where the flow including
  differential rotation is almost unaffected by the magnetic field. In
  the cases with the highest supercriticality the magnetic field is
  permanently strong and differential rotation permanently quenched,
  both having relatively little temporal variation. In an intermediate
  regime both the magnetic fields and differential rotation exhibit
  strong anticorrelated fluctuations.}%
{The occurrence of grand minima in the Sun has been conjectured to be
  due to a weakly supercritical dynamo. The current results support
  this idea and suggest that the absence of grand minima in earlier
  global dynamo simulations is because their large-scale dynamos were
  too supercritical. Furthermore, the current results suggest that
  the effect of the magnetic field on differential rotation is
  negligible if the average rms-field is less than about 2~kG.}%

  \maketitle


\section{Introduction}

The dynamo-driven solar magnetic field shows variations from cycle to
cycle and on various longer timescales \citep[][and references
therein]{Hathaway_2015_LRSP_12_4,Usoskin_2023_LRSP_20_2}. The most
striking example of this is the Maunder minimum during which sunspots
appeared much less frequently than in modern times
\citep[][]{Maunder_1894_KIMS_18_173, Eddy_1976_Science_192_4245}. At
least three other grand minima have been identified using cosmogenic
isotopes from ice cores and tree rings in the past millennium
\citep[e.g., Fig.~18 of][]{Usoskin_2023_LRSP_20_2}, and several other
epochs of low solar activity are apparent in more distant past
\citep[e.g.][]{Solanki_et_al_2004_Nature_431_1084}.

Modeling of long-term variations of solar activity and dynamos in
general has mostly concentrated on mean-field models \cite[][and
references therein]{Karak_2023_LRSP_20_3}. Mean-field models rely on
parametrizations of unresolved physics or turbulent quantities that
are not directly accessible from the interior of the Sun. Therefore
these quantities have to be extracted from analytic or semi-empirical
theories or from 3D simulations \citep[see,
e.g.][]{Brandenburg_Subramanian_2005_PhRp_417_1,
  Brandenburg_et_al_2023_SSRv_219_55, Kapyla_2025_LRSP_22_3}.
Variations in solar activity are on the one hand thought to arise from
fluctuations of dynamo parameters including some form of $\alpha$
effect, turbulent magnetic diffusivity, and the differential rotation
\citep[e.g.][]{Ossendrijver_2000_AA_359_364,Brandenburg_Spiegel_2008_AN_329_351,Choudhuri_Karak_2012_PRL_109_171103,Oelcek_et_al_2019_SoPh_294_99}. On
the other hand, the back-reaction to large-scale flows is another
avenue to long-term modulations of global dynamos
\citep[e.g.][]{Tobias_1996_AA_307_L21, Brooke_et_al_2002_AA_395_1013,
  Bushby_2006_MNRAS_371_772}. While such modeling efforts have been
successful in reproducing grand minima-like behavior, the levels of
fluctuations of the dynamo parameters remain tunable parameters.

Three-dimensional stellar simulations are now routinely used to study
stellar magnetism \citep[e.g.][]{Kapyla_et_al_2023_SSRv_219_58},
although such models still operate in parameter regimes far removed
from reality \citep[e.g.][]{Kupka_Muthsam_2017_LRCA_3_1}. Several such
simulations have produced solar-like equatorward migrating dynamo
solutions \citep[e.g.][]{Kapyla_et_al_2012_ApJL_755_22,
  Augustson_et_al_2015_ApJ_809_149, Strugarek_et_al_2018_ApJ_863_35,
  Kapyla_2022_ApJ_931_L17}, although the rotation rate in the
simulations is typically higher than in the Sun
\citep[e.g.][]{Warnecke_2018_AA_616_72}, and the resulting cycles are
either too short \citep[e.g.][]{Kapyla_et_al_2019_GAFD_113_149} or too
long \citep[e.g.][]{Ghizaru_et_al_2010_ApJL_715_133}. The issues
related to reproducing the solar dynamo are likely related to the
convective conundrum, which is the discrepancy between solar
observations, theoretical models, and simulations
\citep[e.g.][]{OMara_et_al_2016_AdSpR_58_1475,
  Schumacher_Sreenivasan_2020_RvMP_92_041001}. Grand minima are
another feature that current 3D dynamo simulations struggle to
capture. A few models have produced relatively short epochs of lower
activity in the midst of intense and continuous dynamo action; see
\cite{Augustson_et_al_2015_ApJ_809_149} and
\cite{Kapyla_MJ_et_al_2016_AA_589_56}.

A possible explanation of grand minima in the Sun is that the solar
dynamo is weakly supercritical such that sometimes the magnetic field
remains below the critical value for sunspot formation for extended
periods of time. Recent mean-field models suggest that the solar
dynamo is about two times supercritical
\citep[][]{Wavhal_et_al_2025_SoPh_300_21}. Furthermore, flux-transport
models of a wide variety of stars suggest that slowly rotating stars
are more prone to Maunder-minimum-like events, and that rapidly
rotating stars can have irregular cycles but grand minima do not occur
\citep{Vashishth_et_al_2023_MNRAS_522_2601}. Rotation is a key element
for the inductive effects such as the $\alpha$ effect and differential
rotation. The Sun can therefore be thought to rotate just rapidly
enough to keep a sustained dynamo going. Global 3D dynamo simulations
usually have faster rotation than in the Sun to avoid landing in the
anti-solar differential rotation regime
\citep[e.g.][]{Gastine_et_al_2014_MNRAS_438_L76,Kapyla_et_al_2014_AA_570_43}.
Models with the solar rotation rate and luminosity often lead to this
regime where the dynamo solutions are unrepresentative of the Sun
\citep[e.g.][]{Karak_et_al_2015_AA_576_26,Warnecke_2018_AA_616_72,Strugarek_et_al_2018_ApJ_863_35}.
This can be considered as another manifestation of the convective
conundrum.

Although it is likely that in the Sun the conjectured near-criticality
is precisely because of the rotation-dependent inductive effects,
taking this into account in 3D dynamo simulations is difficult due to
the emergence of anti-solar differential rotation. A notable exception
is the study of \cite{Bushby_et_al_2018_AA_612_97}, where the
supercriticality of the dynamo was varied by changing the rotational
influence on the flow, and in which the differential rotation issue
does not occur because of a specific of Cartesian geometry with
aligned rotation and gravity vectors. To avoid these complications, we
instead vary the magnetic diffusivity to control the supercriticality
of our simulations. The rotation rate is chosen to be three times the
solar value such that the simulations land in the solar-like
differential rotation regime. The current study explores a parameter
regime where the global dynamo varies from weakly to moderately
supercritical, with the goal of determining how supercriticality
affects the occurrence of grand minima and intermittency of
large-scale magnetism.

The remainder of the paper is organized as follows: \Sec{sec:model}
describes the numerical model, \Sec{sec:results} summarizes the
results and our conclusions are presented in \Sec{sec:conclusions}.

\section{The model} \label{sec:model}

The simulation set-up is based on that used in
\cite{Kapyla_2023_AA_669_98}. The simulation domain is a spherical
wedge that spans $\rin<r<R$ in radius, where $\rin=0.7R$ and $R$ is
the radius of the star, $\theta_0<\theta<\pi-\theta_0$ in colatitude,
where $\theta_0=\pi/12$, and $0<\phi<\pi/2$ in longitude. Equations of
fully compressible magnetohydrodynamics (MHD) were solved,
\begin{eqnarray}
\frac{\pd {\bm A}}{\pd t} & = & {\bm U} \times {\bm B} - \eta \mu_0 {\bm J}, \label{equ:indu}\\
\frac{\DD \ln\rho}{\DD t} & = & -\bm\nabla\bm\cdot{\bm U}, \label{equ:conti}\\
\frac{\DD {\bm U}}{\DD t} & = &  {\bm g}  - 2\bm\Omega_0\times{\bm U} - \frac{1}{\rho}(\bm\nabla p - {\bm J} \times {\bm B} - \bm\nabla \bm\cdot 2\nu\rho\bm{\mathsf{S}}),\label{equ:NS}\\
\rho T\frac{\DD s}{\DD t} & = & -T\bm\nabla\bm\cdot {\bm F}^{\rm SGS} -\bm\nabla\bm\cdot {\bm F}^{\rm rad} + 2\nu\rho \bm{\mathsf{S}}^2 + \eta\mu_0 {\bm J}^2, \label{equ:ss}
\end{eqnarray}
where ${\bm A}$ is the magnetic vector potential, ${\bm U}$ is the
velocity, ${\bm B} = \bm\nabla\times{\bm A}$ is the magnetic field,
$\eta$ is the magnetic diffusivity, $\mu_0$ is the permeability of
vacuum, ${\bm J}=\bm\nabla\times{\bm B}/\mu_0$ is the current density,
$\DD/\DD t = \pd/\pd t + {\bm U}\bm\cdot\bm\nabla$ is the advective
time derivative, $\rho$ is the density, and ${\bm g}=-\bm\nabla \phi$
is the acceleration due to gravity, where $\phi = -GM/r$ is a fixed
spherically symmetric gravitational potential, with $G$ and $M$ being
the universal gravitational constant and the stellar mass,
respectively. $\bm\Omega_0=\Omega_0(\cos\theta,-\sin\theta,0)^{\rm T}$
is the angular velocity vector, where $\Omega_0$ is the rotation rate
of the frame of reference, $p$ is the pressure, $\nu$ is the kinematic
viscosity, $T$ is the temperature, and $s$ is the specific entropy
with $Ds=\cV D\ln p-\cP D\ln\rho$, where $\cV$ and $\cP$ are the
specific heat capacities at constant volume and pressure,
respectively. The gas obeys the ideal gas law, $p=\mathcal{R}\rho T$,
where $\mathcal{R}=\cP-\cV$ is the gas constant. The rate of strain
tensor is given by
\begin{eqnarray}
\mathsf{S}_{ij}\, = \,\onehalf (U_{i;j} + U_{j;i}) - \onethird \delta_{ij} \bm\nabla\bm\cdot {\bm U},
\end{eqnarray}
where the semicolons refer to covariant derivatives
\citep{Mitra_et_al_2009_ApJ_697_923}. The radiative flux is given by
the diffusion approximation $\FFFrad = - K \bm\nabla T$, where $K$ is
the heat conductivity. Following \cite{Kapyla_2023_AA_669_98}, we use
$K = K_1 + K_2$, where $K_1 = K_1(r)$ is a fixed function of radius
and $K_2 = K_2(\rho,T)$ is computed according to Kramers opacity law
\citep[e.g.][]{Weiss_et_al_2004_Cox_Giuli_Principles}. $K_1$ is given
by
\begin{eqnarray}
K_1 = K_{\rm top} \left[1 + \tanh\left(\frac{r-R}{d_K}\right)\right],
\end{eqnarray}
where $K_{\rm top}=2 \cP\Fbot/|g(R)|$, with $\Fbot=L/4\pi\rin^2$ where
$L$ is the luminosity of the star, and where $d_K = 0.015R$. $K_2$ is
given by
\begin{eqnarray}
K_2(\rho,T)\, =\, K_0 (\rho/\rho_0)^{-2} (T/T_0)^{13/2},
\label{equ:Krad2}
\end{eqnarray}
where $\rho_0$ and $T_0$ are reference values of density and
temperature.

The subgrid scale (SGS) flux is given by
\begin{eqnarray}
  {\bm F}^{\rm SGS}\, = \,-\,\chiSGS \rho \bm\nabla s',
\label{equ:FSGS}
\end{eqnarray}
where $\chiSGS$ is the (constant) SGS diffusion coefficient for the
entropy fluctuation $s'(r,\theta,\phi)=s-\brac{s}_{\theta\phi}$, where
$\brac{s}_{\theta\phi}$ is the spherically symmetric part of the
specific entropy. The SGS flux contributes only negligibly to the net
radial energy transport because it is decoupled from the mean
stratification.

The simulations were run with the {\sc Pencil
  Code}\footnote{\href{https://pencil-code.org/}{https://pencil-code.org/}}
\citep{Pencil_Code_Collaboration_2021_JOSS_6_2807}. The code was
configured to use third-order temporal and sixth-order spatial
discretisation in the current study. Advective terms in
\Equsa{equ:indu}{equ:ss} are written as fifth-order upwinding
derivatives with a sixth-order hyperdiffusive correction, where the
diffusion coefficient is flow-dependent; see Appendix~B of
\cite{Dobler_et_al_2006_ApJ_638_336}.

\subsection{System parameters and diagnostic quantities}
\label{sec:syspar}

The simulations are defined by $\Fbot$, $K_{0}$, $\rho_0$, $T_0$,
$\Omega_0$, $\nu$, $\eta$, and $\chiSGS$. The current models use a
significantly enhanced luminosity in comparison to real stars to bring
the thermal and dynamical timescales close enough to be resolved in
the simulations. This leads to correspondingly higher convective
velocities, and therefore, the rotation rate is increased accordingly
to capture a similar rotational influence on the flow in the
simulations in comparison to real stars; see Appendix~A of
\cite{Kapyla_et_al_2020_GAFD_114_8}.

The non-dimensional luminosity is given by
$\mathcal{L} =L/(\rho_0 (GM)^{3/2}R^{1/2})$, where $\rho_0$ is the
initial density at the base of the convection zone. The degree of
luminosity enhancement is given by the ratio
$L_{\rm ratio}=\mathcal{L}/\mathcal{L}_\odot \approx 2.1\cdot 10^5$,
where $\mathcal{L}_\odot$ is the corresponding dimensionless solar
luminosity. The initial stratification is given by the non-dimensional
pressure scale height at the surface, $\xi_0 = \calR T_1/(GM/R)$,
where $T_1=T(R,t=0)$.

The system parameters include the SGS, thermal, and magnetic Prandtl
numbers and the Taylor number
\begin{eqnarray}
\PraSGS = \frac{\nu}{\chiSGS}, \ \ \Pra = \frac{\nu}{\chi}, \ \ \PrM =\frac{\nu}{\eta}, \ \   \Ta = \frac{4\Omega_0^2 d^4}{\nu^2},
\end{eqnarray}
where $d$ is a global length scale characterizing the system. The
efficiency of convection is quantified by the flux-based Rayleigh
number,
\begin{eqnarray}
  \RaF \,= \,\frac{g_0 d^4 \Fbot}{\cP \rho_{0} T_{0} \nu \chi_{0}^2},
\end{eqnarray}
where $g_0 = GM/\rin^2$, and $T_{0}$ and $\chi_{0}$ are the
temperature and radiative diffusivity in the initial state at
$r=\rin$, respectively. A flux-based diffusion-free modified Rayleigh
number is given by
\citep[e.g.][]{Christensen_2002_JFM_470_115,Christensen_Aubert_2006_GeJI_166_97}
\begin{eqnarray}
\RaFS = \frac{\RaF}{\Pra^2\Ta^{3/2}} = \frac{g_0 \Fbot}{8\cP \rho_{0} T_{0} \Omega_0^3 d^2}.
\end{eqnarray}
Choosing $d \equiv \cP T_{0}/g_0$ and
$\ustar \equiv (\Fbot/\rho_{0})^{1/3}$ yields
\begin{eqnarray}
  \RaFS = \frac{\Fbot}{8\rho_{0}\Omega_0^3 d^3} = \CoF^{-3},
\end{eqnarray}
where $\CoF = 2\Omega_0 d /\ustar$ is a flux-based Coriolis number
\citep[e.g.][]{Kapyla_2024_AA_683_221, Bekki_2025_AA_703_262}.

The fluid and magnetic Reynolds numbers and the P\'eclet number are
given by
\begin{eqnarray}
\Rey = \frac{\urms}{\nu k_1}, \ \  \ReM = \frac{\urms}{\eta k_1}, \ \ \Pe = \frac{\urms}{\chiSGS k_1},
\label{equ:Rey}
\end{eqnarray}
respectively, where
$\urms = \left\langle\sqrt{\threehalfs
    (U_r^2+U_\theta^2)}\right\rangle$ is the time- and volume-averaged
rms velocity, where contributions from differential rotation are
neglected. The inverse of the wavenumber
$k_1=2\pi/\Delta r\approx21/R$ is used to characterize the radial
extent of the convection zone. The total rms-velocity $\Urms$, where
$U_\phi$ is retained, is used to characterize the families of
solutions found in the study. The fluid Coriolis number is given by
\begin{eqnarray}
\Co = \frac{2\Omega_{0}}{\urms k_{1}}.
\end{eqnarray}
Mean quantities are denoted by overbars and are defined as the time-
and azimuthal averages,
\begin{eqnarray}
  \mean{f}(r,\theta) = \frac{1}{\Delta \phi \Delta t} \int_{t_0}^{t_0
    + \Delta t}\!\!\int_0^{\Delta \phi} f(r,\theta,\phi,t) d\phi dt,
\end{eqnarray}
where $t_0$ and $\Delta t$ are the beginning and the length of the
statistically steady part of the simulation, and where $\Delta \phi =
\pi/2$ is the azimuthal extent of the simulation domain. Error
estimates were obtained by dividing the time series into three parts and
by computing averages over each one of them. The largest deviation of
these sub-averages from the average over the whole time series was
taken to represent the error.

\begin{table}[t!]
\centering
\caption[]{Summary of the runs.}
  \label{tab:runs}
  \vspace{-0.5cm}
  $$
  \begin{array}{p{0.08\linewidth}cccccc}
    \hline
    \hline
    \noalign{\smallskip}
    Run  & \PrM  & \Rey  & \ReM  & \Co & \tilde{\lambda} & \mbox{Regime}\\
\hline
 hydro & - & 24.7 &  - & 4.95 & - & -  \\
 A & 0.20 & 24.7 &  4.9 & 4.96 & (-2.51\pm0.64)\cdot10^{-3} & -  \\
 B & 0.33 & 24.7 &  8.2 & 4.96 &  (1.27\pm0.45)\cdot10^{-3} & \mbox{I} \\
 C & 0.50 & 24.6 & 12.3 & 4.97 &  (1.66\pm0.22)\cdot10^{-3} & \mbox{I}  \\
 D & 0.60 & 24.5 & 14.7 & 5.00 &  (1.48\pm0.94)\cdot10^{-3} & \mbox{II} \\
 E & 0.67 & 23.9 & 15.9 & 5.12 &  (1.47\pm0.31)\cdot10^{-3} & \mbox{II} \\
 F & 0.75 & 23.7 & 17.8 & 5.16 &  (1.54\pm0.87)\cdot10^{-3} & \mbox{II} \\
 G & 1.00 & 23.5 & 23.5 & 5.20 &  (6.89\pm4.68)\cdot10^{-3} & \mbox{II} \\
 H & 1.50 & 23.5 & 35.3 & 5.21 &  (2.39\pm0.35)\cdot10^{-2} & \mbox{III} \\
 I & 2.00 & 23.1 & 46.2 & 5.30 &  (4.11\pm1.36)\cdot10^{-2} & \mbox{III} \\
\hline
  \end{array}
  $$
  \tablefoot{$\PraSGS=1$ and grid resolution
    $128 \times 288 \times 144$ was used in all of the
    runs. Furthermore, all runs have $\Ta=8.42 \cdot 10^6$,
    $\RaF=1.25\cdot 10^9$, $\RaFS=3.47\cdot 10^{-5}$, and
    $\CoF=30.7$. The penultimate column shows the kinematic growth
    rate of the magnetic field $\lambda$ and its error, normalized by
    the inverse convective turnover time $\urms k_{1}$. The last
    column indicates the dynamo and flow regime of the simulation as
    defined in \Sec{subsec:timevo}.}
\end{table}

\subsection{Initial and boundary conditions}

Initially, the stratification is isentropic with a polytropic index
$n=\nad=1.5$ and $\xi_0=0.02$, resulting in an initial density
contrast of $30$. The value of $K_0$ is chosen such that
$F_{\rm rad}=F_{\rm tot}$ at the bottom of the domain. The velocity
and magnetic fields were initialized with random Gaussian noise
fluctuations with amplitudes $0.4$~$\mbox{m/s}$ and $6.5$~G,
respectively.

The radial and latitudinal boundaries are assumed to be impenetrable
and stress-free for the flow. At the bottom boundary, a fixed heat
flux is prescribed, while at the top, a black-body condition is
applied. At the latitudinal boundaries, the gradients of thermodynamic
quantities are set to zero. For the magnetic field, we apply a radial
field condition at the upper and a perfect conductor condition at the
lower boundary. At the latitudinal boundaries, the field is assumed to
be tangential to the boundary. The azimuthal direction is periodic for
all quantities. The boundary conditions are described in detail in
Section~2.1 of \cite{Kapyla_et_al_2013_ApJ_778_41}.

\begin{figure*}
  \begin{center}
  \includegraphics[width=\textwidth]{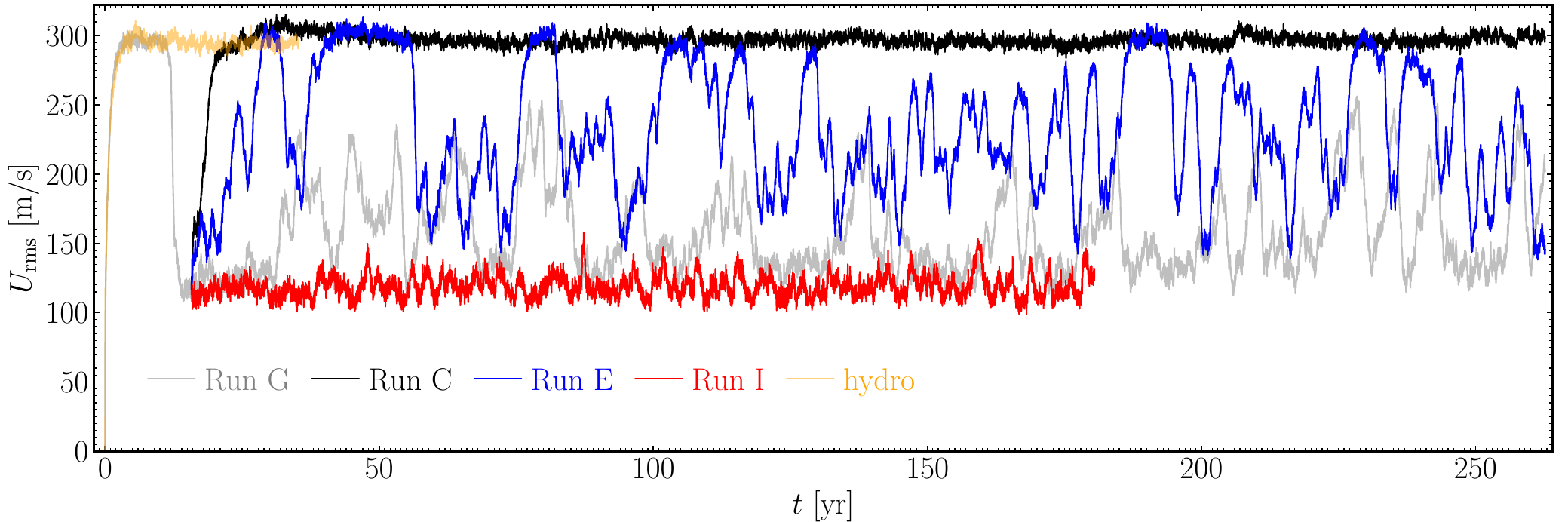}
  \includegraphics[width=\textwidth]{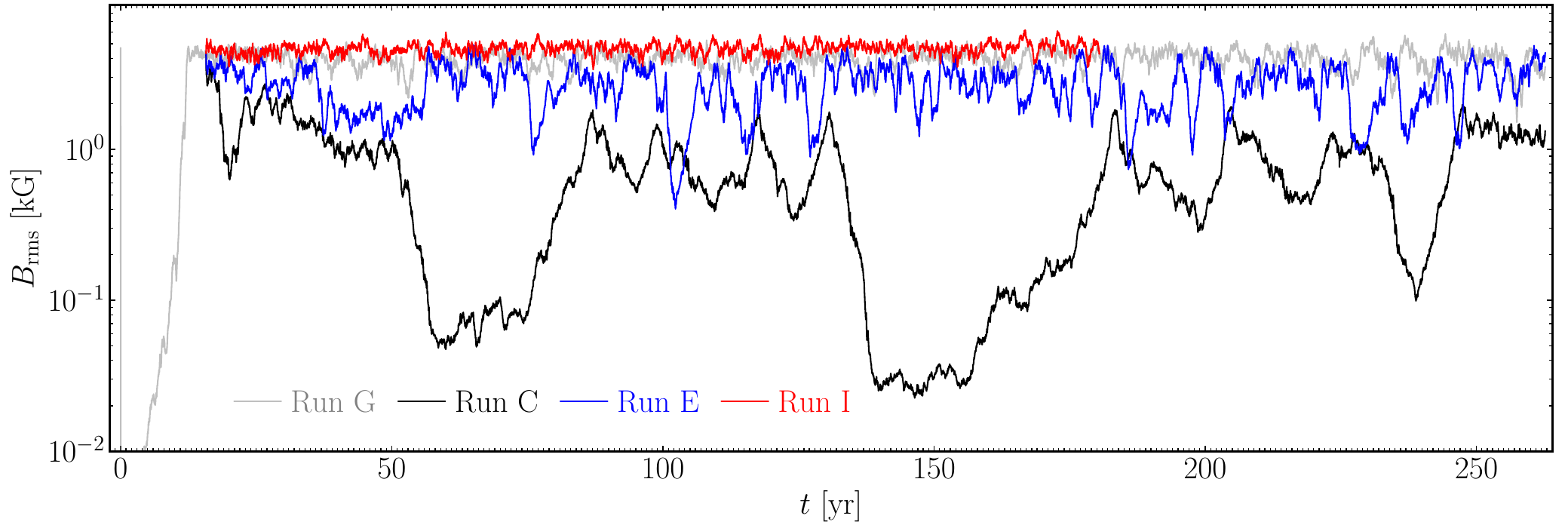}
  \end{center}
  \vspace{-.5cm}
  \caption{Total rms velocity (top) and magnetic field (bottom) as
    functions of time for Runs~C (black, $\PrM=0.5$), E (blue,
    $\PrM = 0.666$), and I (red, $\PrM = 2$). The fainter grey and
    orange lines show the progenitor run (Run~G) and the hydrodynamic
    run (hydro), respectively.}
\label{fig:ptsuB3}
\end{figure*}

\section{Results}
\label{sec:results}

The main runs of the present study consist of nine models with $\PrM$
ranging from 0.2 to 2; see Runs~A to I in \Table{tab:runs}. One of
these runs (Run~G with $\PrM=1$) was started from the initial
conditions discussed above whereas the rest of the main runs were
branched from a snapshot of Run~G at $t = 15.8$~yr; see
\Figa{fig:ptsuB3}. All of the main runs except Run~I were integrated
until $t\approx 264$~yr. Run~I is shorter due to a numerical
instability that develops around $t=180$~yr. This is likely due to the
fact that the $\PrM=2$ case is near the resolving limit of the current
resolution. The hydrodynamic (hydro) case was run to $t\approx
40$~yr. Kinematic growth rates of the magnetic field (sixth column of
\Table{tab:runs}) were obtained from separate runs for each $\PrM$
which were started from a snapshot of Run~A without a dynamo where the
magnetic field had decayed to a value of the order of $10^{-3}$~G.

Three non-rotating MHD runs with $\PrM=1$, $1.5$, and $2$ were
additionally made (not listed). None of the non-rotating runs have
dynamos suggesting that a small-scale dynamo is absent in the current
simulations. Runs~C, E, and I with $\PrM = 0.5$, $0.666$, and $2$,
respectively, are discussed as representative cases for the three main
regimes of solutions found in the present study.

\subsection{Time evolution of integral quantities}
\label{subsec:timevo}

\Fig{fig:ptsuB3} shows the time series of the total rms velocity
$\Urms$ and rms magnetic field $\Brms$ for Runs~C, E, and I. These
runs are representative of the low, intermediate, and high states of
the dynamo. The velocity field of Run~C is practically unaffected by
the magnetic field and it is very similar to the hydrodynamic
case. The total rms magnetic field in this run fluctuates between
around 30~G and 2~kG; see the lower panel of \Figa{fig:ptsuB3}. The
timescale of these fluctuations is of the order of decades but the
field is always weak enough such that the rms velocity remains
practically unaffected at all times. In Run~E the rms velocity shows
strong fluctuations, ranging from a minimum of around 150~m/s to
occasional phases during which it reaches the hydrodynamic value of
nearly 300~m/s. This is associated with a magnetic field that is
consistently stronger than in runs with lower $\PrM$. In Run~E the rms
magnetic field has several excursions below 2~kG which is still
sufficiently weak not to affect the velocity field
appreciably. However, most of the time the magnetic field is stronger
than this and has a significant influence on the flow. The rms
magnetic field typically fluctuates between roughly 2 to 4~kG on a
timescale of a few years, although there are also longer periods of
weak field, for example, between 40-55~yrs and 185-195~yrs. Finally,
in Run~I the flow is quenched to an average of 120~m/s throughout the
simulation and the rms magnetic field is around 4.5~kG with much less
variation than in the lower $\PrM$ cases.

\begin{figure*}
  \begin{center}
  \includegraphics[width=\textwidth]{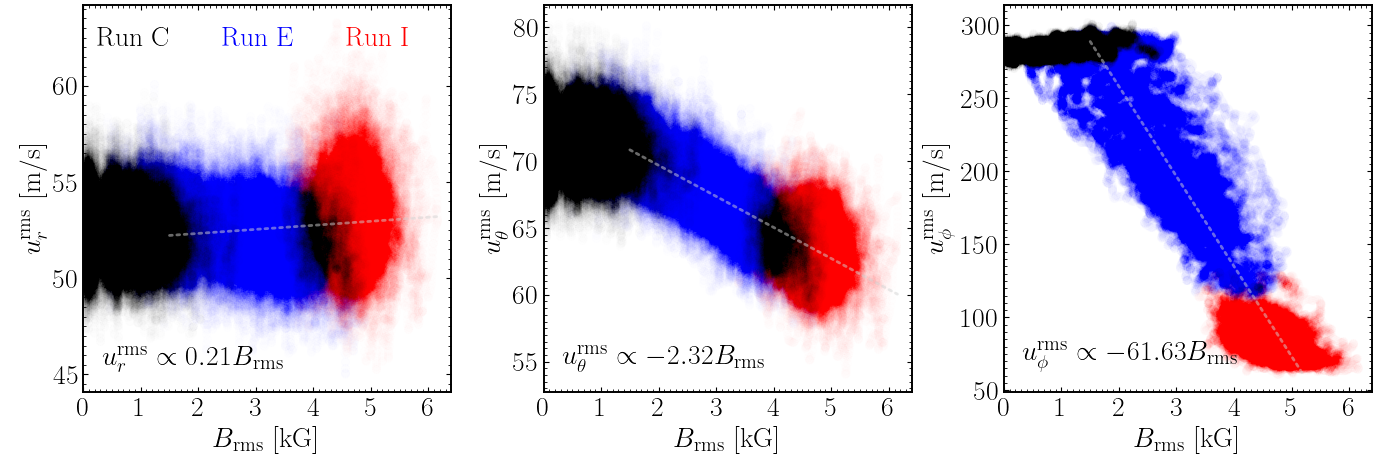}
  \end{center}
  \vspace{-.5cm}
  \caption{Scatter plots of rms values of the radial (left),
    latitudinal (middle), and azimuthal (right) rms velocities as
    functions of $\Brms$ for Runs~C, E, and I. The grey dashed lines
    show linear fits to the data for $\Brms > 1.5$~kG, and the fitted
    formulae are indicated by the legends. The constant of
    proportionality has the unit ${\rm (m/s)/kG}$.}
\label{fig:pcorruB}
\end{figure*}

Runs~C, E, and I are prototypical of the three distinct classes of
solutions that can be identified:
\begin{itemize}
\item Regime I (weak supercriticality): the flow is practically
  unaffected by the magnetic field and it exhibits weak
  fluctuations. The magnetic fields have strong fluctuations on
  timescales of decades.
\item Regime II (intermediate supercriticality): both the flow and
  the magnetic field vary in anticorrelation between high and low
  states with strong fluctuations in both. Timescale of the
  fluctuations varies between a few years to roughly a decade.
\item Regime III (high supercriticality): the magnetic field is
  permanently strong and differential rotation permanently quenched
  with weak fluctuations of both.
\end{itemize}
Long periods of low magnetic activity appear in cases where the dynamo
is the least supercritical corresponding to Regime~I. Such epochs are
reminiscent of grand minima of stellar magnetic activity, the most
well-known being the Maunder minimum of the Sun
\citep[e.g.][]{Usoskin_2023_LRSP_20_2}.

\begin{figure*}
  \begin{center}
  \includegraphics[width=0.5\textwidth]{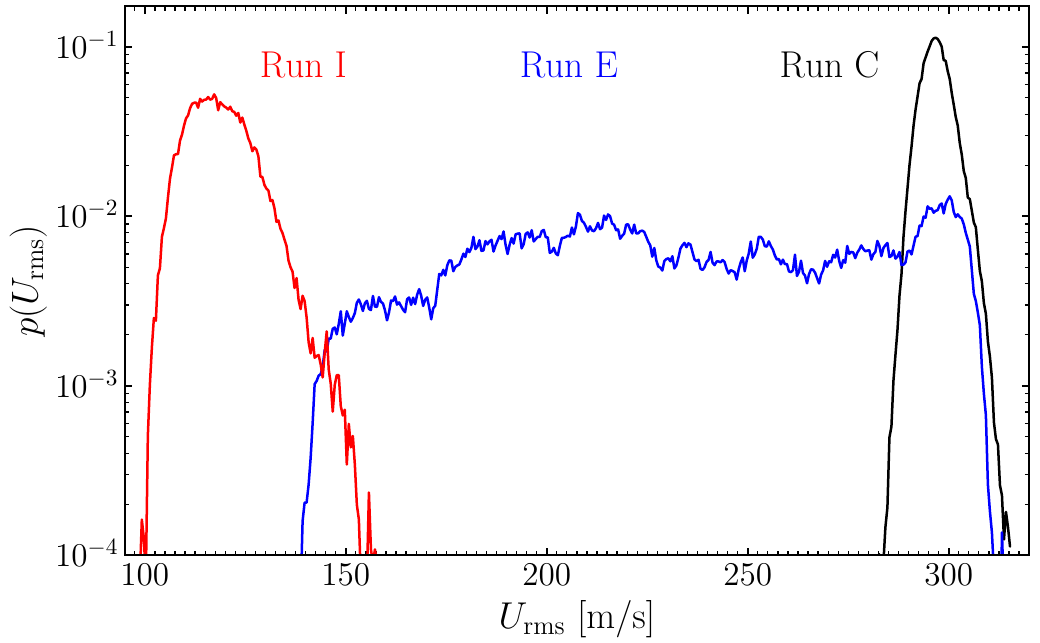}\includegraphics[width=0.5\textwidth]{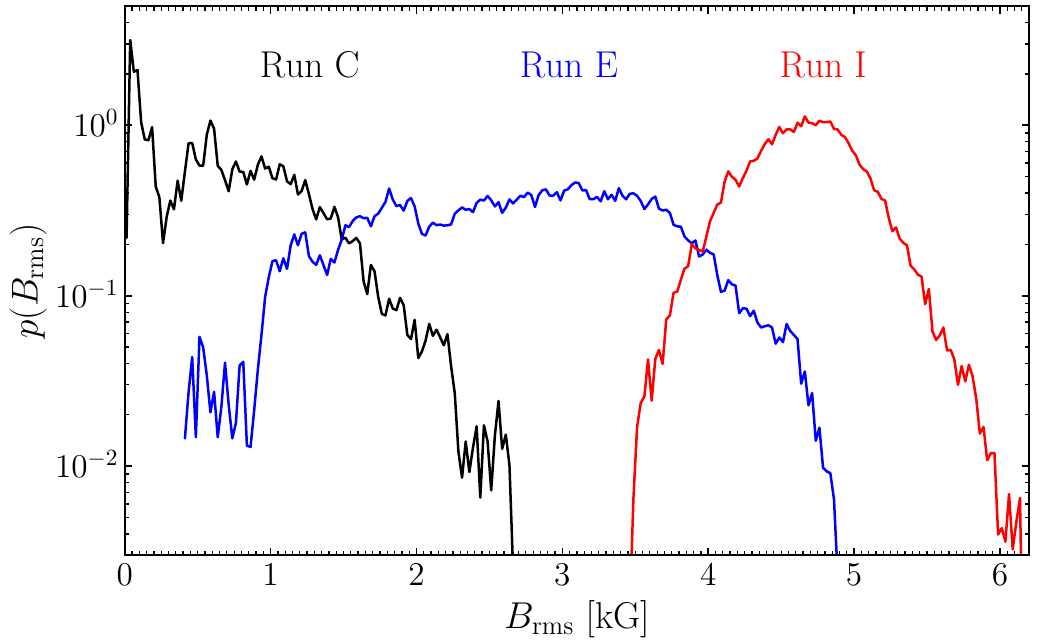}
  \end{center}
  \vspace{-.5cm}
  \caption{Probability density functions of $\Urms$ and $\Brms$ from
    the same runs as in \Fig{fig:pcorruB}.}
\label{fig:phistuB}
\end{figure*}

\begin{table*}[t!]
\centering
\caption[]{Summary of diagnostics.}
\label{tab:statistics}
\vspace{-0.5cm}
$$
\begin{array}{p{0.03\linewidth}cccccccccccccc}
  \hline
  \hline
  \noalign{\smallskip}
  Run & \Urms^{\rm mean} & \Urms^{\rm median} & \Urms^{\rm min} & \Urms^{\rm max}  & \Urms^{\rm std} & \mathcal{S}_U & \mathcal{K}_U & \Brms^{\rm mean} & \Brms^{\rm median} & \Brms^{\rm min} & \Brms^{\rm max}  & \Brms^{\rm std} & \mathcal{S}_B & \mathcal{K}_B \\
  \hline
A & 294 & 294 & 284 & 304 &  3.0 &  0.08 &  2.98 &   - & - & - & - & - & - & - \\
B & 295 & 295 & 283 & 308 &  3.3 &  0.07 &  2.91 & 0.34 & 0.29 & 0.05 & 1.20 & 0.26 &  1.21 & 3.94 \\
C & 297 & 297 & 284 & 315 &  3.8 &  0.48 &  3.56 & 0.66 & 0.59 & 0.02 & 2.71 & 0.52 &  0.66 & 2.91 \\
D & 290 & 301 & 130 & 318 & 33.2 & -2.93 & 11.10 & 1.81 & 1.79 & 0.14 & 5.01 & 0.74 &  0.59 & 5.30 \\
E & 235 & 231 & 138 & 314 & 44.9 & -0.03 &  1.86 & 2.66 & 2.73 & 0.40 & 4.92 & 0.90 & -0.10 & 2.26 \\
F & 212 & 209 & 112 & 315 & 45.5 &  0.29 &  2.37 & 3.15 & 3.18 & 1.18 & 5.20 & 0.75 & -0.08 & 2.35 \\
G & 159 & 149 & 111 & 256 & 31.2 &  0.92 &  3.01 & 3.98 & 4.04 & 1.51 & 5.81 & 0.56 & -0.51 & 3.52 \\
H & 125 & 124 & 102 & 182 & 10.0 &  1.24 &  5.67 & 4.44 & 4.44 & 2.84 & 6.00 & 0.43 & -0.12 & 3.31 \\
I & 119 & 118 &  99 & 158 &  8.1 &  0.69 &  3.74 & 4.66 & 4.67 & 3.37 & 6.15 & 0.40 &  0.04 & 3.14 \\
  \hline
\end{array}
$$
\tablefoot{The mean, median, minimum (min), maximum (max), standard
  deviation (std), skewness ($\calS$), and kurtosis ($\calK$) of
  $\Urms$ and $\Brms$ are listed. Velocity (magnetic field) is given
  in units of m/s (kG). Magnetic quantities are not listed from Run~A
  because it has no dynamo.}
\end{table*}

\Fig{fig:pcorruB} shows scatter plots of rms values of different
velocity components from the three representative runs. The radial
velocity ($u^{\rm rms}_{r}$) is practically unaffected by the magnetic
field in all cases, although a minor increase is seen for Run~I. The
latitudinal rms velocity ($u^{\rm rms}_{\theta}$) decreases by about
10~m/s when the magnetic field increases from one to five kilogauss,
corresponding to a relative change of about 15 per cent. The decrease
is approximately linear as a function of $\Brms$ above
$\Brms\approx 1.5$~kG. The azimuthal velocity ($u^{\rm rms}_{\phi}$)
is by far the most affected component, as seen in the right panel of
\Figa{fig:pcorruB}. The rms-value decreases from around 280~m/s in the
low-$\Brms$ quasi-hydrodynamic state in Run~C to less than 100~m/s in
the strongly quenched state of Run~I. As in the case of the
latitudinal velocity, the decrease is roughly linear in the magnetic
field strength. The azimuthal rms velocity is dominated by the
axissymmetric component, that is, differential rotation. Therefore,
the main effect of the magnetic field is to quench the differential
rotation. The current data gives a threshold value of around
$\Brms\approx 2$~kG above which the average magnetic field starts to
have a significant effect on the large-scale flows. At face value, the
current results suggest that the average magnetic field in the solar
convection zone should be less than about 2~kG for the strong
differential rotation to be maintained. However, this preliminary
conjecture should be taken with caution because the current
simulations have shortcomings that are discussed further in
\Sec{sec:relevance}.

\begin{table*}[t!]
\centering
\caption[]{Fractions of time that $\Urms$ and $\Brms$ spent in various
  states above and below their respective temporal averages.}
\label{tab:velmaglohi}
\vspace{-0.5cm}
$$
\begin{array}{p{0.05\linewidth}rrrrrrrrrrrrrrrrrrrr}
  \hline
  \hline
  \noalign{\smallskip}
  Run & f^{\Urms}_{-0.02} & f^{\Urms}_{-0.05} & f^{\Urms}_{-0.10} & f^{\Urms}_{-0.20} & f^{\Urms}_{-0.50} & f^{\Urms}_{+0.02} & f^{\Urms}_{+0.05} & f^{\Urms}_{+0.10} & f^{\Urms}_{+0.20} & f^{\Urms}_{+0.50} & f^{\Brms}_{-0.02} & f^{\Brms}_{-0.05} & f^{\Brms}_{-0.10} & f^{\Brms}_{-0.20} & f^{\Brms}_{-0.50} & f^{\Brms}_{+0.02} & f^{\Brms}_{+0.05} & f^{\Brms}_{+0.10} & f^{\Brms}_{+0.20} & f^{\Brms}_{+0.50}\\
\hline
A & 0.03 & 0.00 & 0.00 & 0.00 & 0.00 & 0.02 & 0.00 & 0.00 & 0.00 & 0.00 & - & - & - & - & - & - & - & - & - & - \\
B & 0.03 & 0.00 & 0.00 & 0.00 & 0.00 & 0.03 & 0.00 & 0.00 & 0.00 & 0.00 & 0.46 & 0.45 & 0.44 & 0.40 & 0.28 & 0.44 & 0.43 & 0.40 & 0.33 & 0.22 \\
C & 0.04 & 0.00 & 0.00 & 0.00 & 0.00 & 0.03 & 0.00 & 0.00 & 0.00 & 0.00 & 0.50 & 0.48 & 0.46 & 0.41 & 0.28 & 0.41 & 0.40 & 0.38 & 0.35 & 0.26 \\
D & 0.14 & 0.13 & 0.11 & 0.07 & 0.01 & 0.66 & 0.22 & 0.00 & 0.00 & 0.00 & 0.50 & 0.46 & 0.40 & 0.27 & 0.15 & 0.39 & 0.37 & 0.33 & 0.23 & 0.08 \\
E & 0.48 & 0.43 & 0.34 & 0.17 & 0.00 & 0.39 & 0.35 & 0.28 & 0.14 & 0.00 & 0.40 & 0.38 & 0.33 & 0.26 & 0.08 & 0.49 & 0.46 & 0.41 & 0.31 & 0.06 \\
F & 0.41 & 0.36 & 0.29 & 0.18 & 0.00 & 0.45 & 0.40 & 0.32 & 0.21 & 0.00 & 0.46 & 0.43 & 0.37 & 0.23 & 0.01 & 0.41 & 0.37 & 0.31 & 0.19 & 0.01 \\
G & 0.49 & 0.45 & 0.36 & 0.09 & 0.00 & 0.28 & 0.25 & 0.20 & 0.14 & 0.02 & 0.31 & 0.26 & 0.18 & 0.08 & 0.00 & 0.41 & 0.33 & 0.22 & 0.06 & 0.00 \\
H & 0.40 & 0.24 & 0.04 & 0.00 & 0.00 & 0.32 & 0.22 & 0.12 & 0.03 & 0.00 & 0.38 & 0.27 & 0.14 & 0.02 & 0.00 & 0.38 & 0.27 & 0.13 & 0.01 & 0.00 \\
I & 0.37 & 0.22 & 0.04 & 0.00 & 0.00 & 0.33 & 0.20 & 0.08 & 0.01 & 0.00 & 0.32 & 0.22 & 0.11 & 0.01 & 0.00 & 0.39 & 0.26 & 0.12 & 0.02 & 0.00 \\
\hline
\end{array}
$$
\end{table*}

\subsection{Statistics of activity}

Probability density functions (PDFs) of $\Urms$ and $\Brms$ are shown
in \Figa{fig:phistuB} for the three representative runs, and
\Table{tab:statistics} lists statistical properties including rms,
median, minimum, maximum, standard deviation, skewness ($\calS$) and
kurtosis ($\calK$) of $\Urms$ and $\Brms$ for all main runs. As can be
expected from the time series, the PDFs of $\Urms$ for Runs~C and I
are relatively narrowly distributed around their respective mean
values, whereas in Run~E the distribution is much wider and
flatter. This is reflected by the standard deviation of $\Urms$ which
is significantly larger for Run~E; see the sixth column of
\Table{tab:statistics}. Furthermore, the skewness and kurtosis of
$\Urms$ in Runs~C and I indicate distributions that are not far from
Gaussian, that is $\calS=0$ and $\calK=3$. In Run~E, $\calS_{U}$ is
nearly zero ($-0.03$) and $\calK_{U}=1.86$, which are reminiscent of a
uniform distribution. The most extreme case in view of skewness and
kurtosis is Run~D, where $\calS_{U}=-2.93$ and $\calK_{U}=11.10$. In
this run the magnetic field is most of the time weak but grows on a
few occasions above the threshold where it strongly affects the flow.

The PDF of $\Brms$ in Run~I is nearly Gaussian, whereas in Run~E the
distribution of $\Brms$ is quite symmetric but much wider, resembling
the corresponding PDF of $\Urms$. The heavily tailed PDF of $\Brms$ in
Run~C reflects the fact that the magnetic field stays substantial
periods of time in states where it is below 300~G, corresponding to
the grand minima-like periods in the simulation.
\Table{tab:statistics} shows that the time-averaged rms magnetic field
increases monotonically with $\PrM$; see the ninth
column. Furthermore, the near-critical cases (Runs~B, C, and D) are
characterized by relatively large positive skewness such that the
magnetic field predominantly resides in a low state with occasional
excursions to higher values. Runs~E and F have small negative
$\calS_{B}$ and $\calK_{B}$ around 2.3, corresponding to relatively
frequent switches of state from a high to a low state with a flat
distribution. As the supercriticality of the dynamo increases further,
the statistics approach a Gaussian distribution (Runs~H and I).

\Table{tab:velmaglohi} lists the fractions of time for which $\Urms$
and $\Brms$ are below or above their respective time-averaged values
by different fractions in each simulation. This is quantified by
$f^{\Urms,\Brms}_{i}$ where $i$ denotes the relative deviation from
the time-averaged value. For example, $f^{\Urms}_{+0.10}$ refers to the
fraction of time that $\Urms$ spent in a state where its value is 10
per cent above the time average over the statistically steady part of the
simulation\footnote{Minus the initial transient.}, and
$f^{\Brms}_{-0.05}$ gives the fraction of time the simulation spent in
a state where $\Brms$ was five per cent below the time-averaged
value. These quantities can be used to further characterize the
intermittency and fluctuations in the simulations. In the cases
corresponding to Regime~I (Runs~B and C), the fluctuations of $\Urms$
are always less than five per cent from the corresponding temporal
averages (see the third to eighth columns of \Table{tab:velmaglohi}),
whereas the magnetic fields show fluctuations of over 50 per cent for
more than 20\% of the time (16th and 21st columns of
\Table{tab:velmaglohi}). In Regime~II (Runs~D, E, F, and G), $\Urms$
has fluctuations greater than 20 per cent for a substantial fraction
(about ten per cent of the time) of the simulations. Fluctuations of
above 50\% of $\Brms$ to either direction are significantly less
frequent in comparison to Regime~I and continue to decrease as the
supercriticality increases. Finally, in Regime~III the fluctuations of
both $\Urms$ and $\Brms$ decrease such that deviation of more than
20\% occur only rarely, typically one to two per cent of the
time. However, the velocity fluctuations remain higher than in
Regime~I even in the most supercritical case (Run~I).

\begin{figure*}
  \begin{center}
  \includegraphics[width=\textwidth]{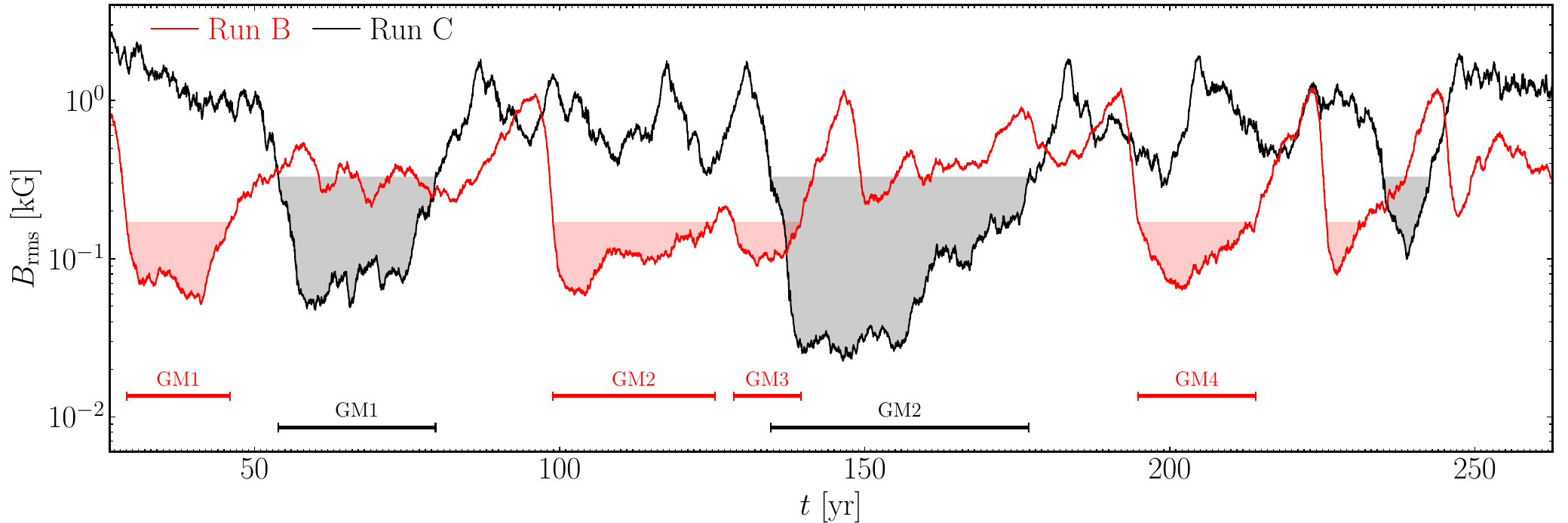}
  \end{center}
  \vspace{-.5cm}
  \caption{Time series of $\Brms$ for Runs~B (red) and C (black). The
    shaded areas indicate periods of time when $\Brms$ is less than
    50\% of its time-averaged value. Periods longer than 10 years,
    corresponding to grand minima (GM) candidates, are indicated by
    the horizontal bars and labels below the data.}
\label{fig:ptsB3_gm}
\end{figure*}

A quantitative definition of a grand minimum applicable to the current
simulations appears to be lacking in the literature. Following
\cite{Usoskin_2023_LRSP_20_2} we refer to a grand minimum as a
prolonged episode (decades or longer) of strongly reduced magnetic
activity which is a manifestation of a low-activity dynamo mode rather
than fluctuations of normal cycles or activity level. As an
illustrative example, we use $f^{\Brms}_{-0.50}$ in the present case
to determine whether the simulations exhibits grand
minima. \Table{tab:gmcandidates} summarizes the results for the
current runs with dynamos. The low activity periods in Runs~B and C
are typically longer than just a few years and can extend to several
decades; see \Fig{fig:ptsB3_gm}. Run~D has a single episode longer
than ten years whereas the rest of the simulations never show
low-activity periods longer than five years. We therefore conclude
that Runs~B and C exhibit grand minima and are characteristic of
Regime~I. The current simulations show such behavior repeatedly which
is distinct from earlier numerical studies
\citep[e.g.][]{Augustson_et_al_2015_ApJ_809_149,Kapyla_MJ_et_al_2016_AA_589_56}
where shorter low activity periods were labeled as grand minima.

\begin{table}[t!]
\centering
\caption[]{Statistics of grand minima candidates.}
\label{tab:gmcandidates}
\vspace{-0.5cm}
$$
\begin{array}{p{0.05\linewidth}rrrrrrrr}
  \hline
  \hline
  \noalign{\smallskip}
  Run & N_{\rm GM} & \langle\Delta t\rangle & \Delta t_{\rm med} & \Delta t_{\rm min} & \Delta t_{\rm max} & N_{1} & N_{5} & N_{10} \\
  \hline
  B &  9 & 8.90 & 5.40 & 0.02 & 26.53 & 5 & 5 & 4 \\
  C & 11 & 7.09 & 0.14 & 0.00 & 42.34 & 4 & 3 & 2 \\
  D &  4 & 5.72 & 5.05 & 0.03 & 12.76 & 3 & 2 & 1 \\
  E & 37 & 0.53 & 0.12 & 0.00 &  3.97 & 7 & 0 & 0 \\
  F & 25 & 0.14 & 0.03 & 0.00 &  0.78 & 0 & 0 & 0 \\
  G &  5 & 0.08 & 0.03 & 0.02 &  0.27 & 0 & 0 & 0 \\
  H &  0 &   -  &   -  &   -  &   -   & 0 & 0 & 0 \\
  I &  0 &   -  &   -  &   -  &   -   & 0 & 0 & 0 \\
  \hline
\end{array}
$$
\tablefoot{$N_{\rm GM}$ is the number of grand minima candidates,
  $\brac{\Delta t}$, $\Delta t_{\rm med}$, $\Delta t_{\rm min}$,
  $\Delta t_{\rm max}$ are the average, median, minimum, and maximum
  duration of grand minima candidates, respectively. $N_{1,5,10}$
  indicate the number of episodes with durations longer than 1, 5, and
  10 years, respectively.}
\end{table}

\begin{figure}
  \begin{center}
\includegraphics[width=0.5\columnwidth]{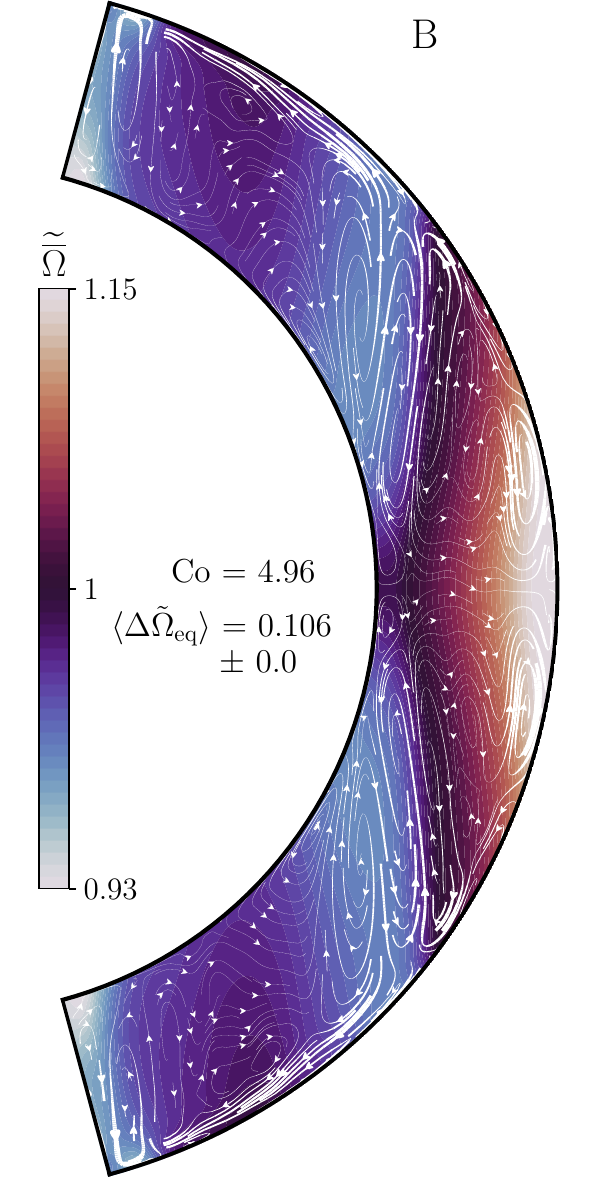}\includegraphics[width=0.5\columnwidth]{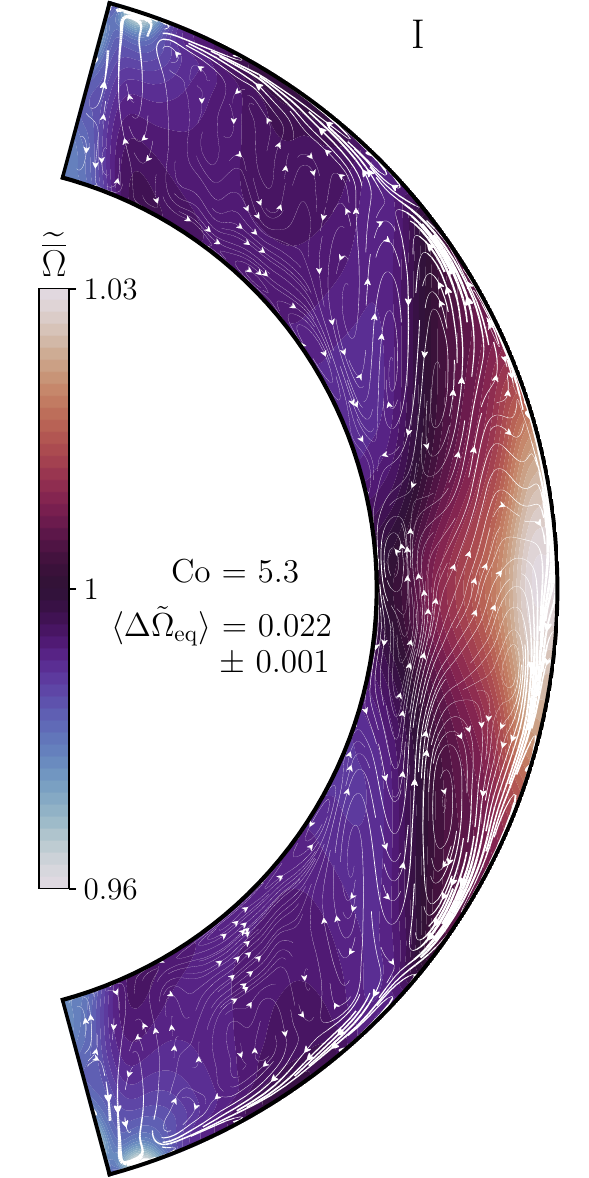}
  \end{center}
  \caption{Colour contours: Time-averaged rotation profiles from
    Runs~B (left) and I (right). The arrows indicate meridional
    flows.}
\label{fig:pOm}
\end{figure}

\begin{figure*}
  \begin{center}
    \includegraphics[width=0.8\textwidth]{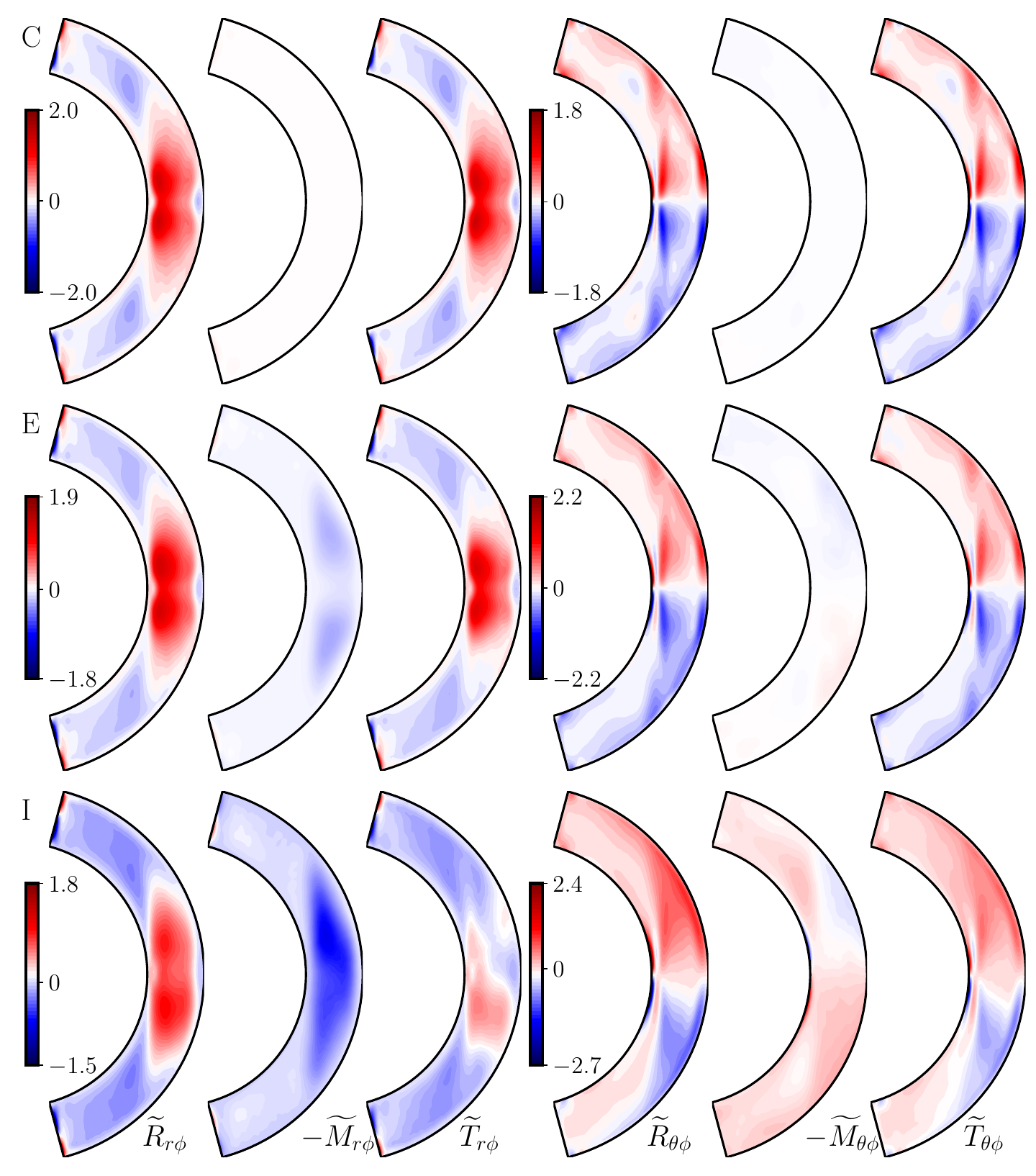}
  \end{center}
  \caption{From left to right: $\Rrp$, $-\Mrp$, $\Trp$, $\Rtp$,
    $-\Mtp$, and $\Ttp$ from Runs~C (top row), E (middle), and I
    (bottom). The tildes refers to normalization by
    $10^{-5}\rho_{0} R^{2}\Omega_{0}^{2}$.}
\label{fig:pfluxcomp}
\end{figure*}

\subsection{Differential rotation}
\label{sec:diffrot}

Time-averaged rotation profiles from Runs~B and I are shown in
\Fig{fig:pOm}. In Run~B the magnetic field remains sufficiently weak
such that the differential rotation is essentially unchanged from the
hydrodynamic case. The large-scale flows are symmetric with respect to
the equator and the maximum amplitude of differential rotation is
$\Delta\Omega/\Omega_{0} = (\Omega_{\rm max} - \Omega_{\rm
  min})/\Omega_{0} = 0.23$. In Run~I the magnetic field exerts a
strong influence on the flow at all times and the differential
rotation is significantly quenched. The amplitude is reduced to
$\Delta\Omega/\Omega_{0} = 0.07$, and the average equatorial
differential rotation, measured as
\citep[e.g.][]{Kapyla_2023_AA_669_98}:
$\brac{\Delta \tilde{\Omega}_{\rm eq}} = \int_{r_{\rm in}}^{R} r^2
[\tilde{\mOm}(r,\theta_{\rm eq}) - 1] dr/(\int_{r_{\rm in}}^{R} r^2
dr)$, decreases by almost a factor of five from $0.106$ to $0.022$ in
comparison to Run~B or the hydrodynamic case. In Run~I the
differential rotation and meridional circulation patterns are
distinctly asymmetric with respect to the equator. This is due to a
hemispheric large-scale dynamo mode excited in the current runs with
sufficiently high $\ReM$; see \Sec{sec:dynsol}.

To see the effects of magnetic field on the angular momentum balance,
we consider the Reynolds and Maxwell stresses associated with the
turbulent (non-axisymmetric) flows and magnetic fields, respectively:
\begin{eqnarray}
\Rij = \mean{\rho u_{i}u_{j}},\ \ \ \Mij = \mu_{0}^{-1}\mean{b_{i}b_{j}},
\end{eqnarray}
where $u_{i} = U_{i}-\mUUi$ and $b_{i} = B_{i}-\mBBi$ and the bars
refer to azimuthal averaging. Time-averaging was applied to $\Rij$ and
$\Mij$ after they were calculated for the full datasets to avoid
including time-variability of the large-scale flows and fields into
the Reynolds and Maxwell stresses. We also quote the total stress
given by $\tij = \Rij - \Mij$, and concentrate on the $r\phi$ and
$\theta\phi$ components that are directly relevant for the angular
momentum transport. \Fig{fig:pfluxcomp} shows the contributions from
Reynolds and Maxwell stresses as well as the total turbulent angular
momentum flux. In the near-critical runs the Maxwell stress is
negligible compared to the Reynolds stress; see the top row of
\Fig{fig:pfluxcomp} for results from Run~C. In Run~E the Maxwell
stress is no longer negligible but the total stress is still dominated
by the hydrodynamic contribution. In Run~I the Maxwell stress reaches
values that are in places comparable to the Reynolds stress especially
for the $r\phi$ component.

The contributions from small-scale Reynolds and Maxwell stresses tend
to have opposite signs which directly translates to weaker driving of
differential rotation. This is in accordance with earlier studies
\citep[e.g.][]{Kapyla_et_al_2017_AA_599_4,Kapyla_2019_AA_622_195,
  Warnecke_et_al_2025_AA_696_93}. In recent high-resolution implicit
large-eddy simulations the small-scale Maxwell stress plays a dominant
role \citep{Hotta_et_al_2022_ApJ_933_199, Hotta_2025_ApJ_985_163}, and
leads to qualitative changes of the differential rotation profile
especially at slow rotation
\citep[][]{Hotta_Hatta_2026_NatAst_10_695}. Such a regime is not
achieved in the current low-resolution cases.

\subsection{Supercriticality of the dynamo}
\label{sec:supercrit}

\begin{figure}[t!]
  \includegraphics[width=\columnwidth]{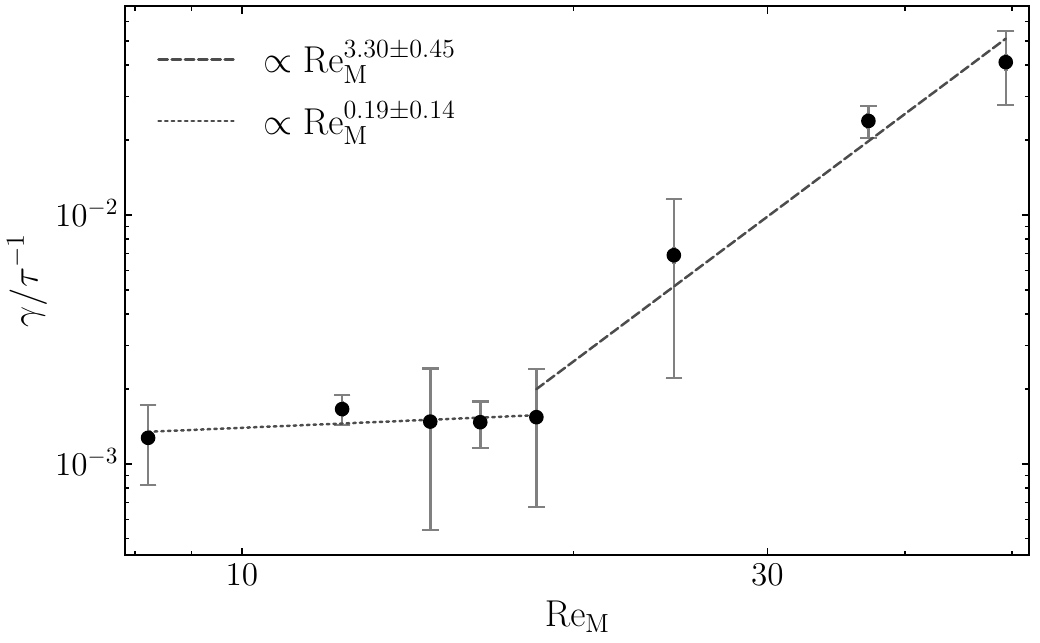}
  \caption{Growth rate of the magnetic field as a function of $\ReM$
    for runs with $\lambda >0$ normalized by the inverse of the
    convective turnover time $\urms k_{1}$.}
\label{fig:pgrowth}
\end{figure}

We used two methods to assess the supercriticality of the
dynamos. First, the critical $\ReM$ lies between 4.9 and 8.2, the
former leading to decay (Run~A), and the latter to a sustained dynamo;
see \Table{tab:runs}. Linear interpolation using the kinematic growth
rates, $\lambda=\pd\ln\Brms/\pd t$, leads to a critical value of
$\Rey_{\rm MC} \approx 7.1$. However, the kinematic growth rates of
the magnetic field behave rather non-linearly for greater values of
$\ReM$\footnote{Kinematic values of $\ReM$ were used in this
  section. In Regime~I runs $\ReM$ is practically the same as in the
  non-linear regime. In Regime~III runs the kinematic $\ReM$ is up to
  roughly seven per cent higher than in the non-linear regime.}; see
\Fig{fig:pgrowth} for the data for runs with $\lambda > 0$. These
cases are characterized by two regimes where the kinematic growth rate
is roughly constant up to $\ReM\approx 20$ with a steep increase for
higher $\ReM$. The supercriticality in terms of $\ReM$,
\begin{eqnarray}
R_{\rm crit} = \frac{\ReM}{\Rey_{\rm MC}},\label{equ:Rcrit}
\end{eqnarray}
varies between 0.7 to 6.9 in the current runs; see the first row of
\Table{tab:supercrit}. The nearly flat part of $\lambda$ for
$\ReM\lesssim 20$ is possibly because the relevant turbulent dynamo
parameters are already sufficiently independent of
$\ReM$. Furthermore, the excited dynamo mode does not change in this
regime: in all of the kinematic runs with $\ReM\lesssim 20$, a
high-latitude quasi-stationary dynamo mode dominates (not shown). The
dominant dynamo mode in the kinematic regime changes around
$\ReM \approx 20$ such that for Runs~G, H, and I a poleward migrating
cyclic mode near the equator dominates, whereas in Run~F the
high-latitude and near-equator modes appear to coexist with roughly
equal strength.

In the second approach the $\alpha\Omega$ dynamo was taken to
correspond to the dynamos in the simulations. While this is possible,
the plausibility of this assumption has not been checked in detail,
for example, with test field calculations and associated mean-field
modeling \citep[e.g.][]{Warnecke_et_al_2021_ApJL_919_13}. In the
$\alpha\Omega$ case the supercriticality of the dynamo depends on the
dynamo numbers
\begin{eqnarray}
C_\alpha = \frac{\alpha \Delta R}{\etaT},\ \ \ C_\Omega = \frac{\Delta \Omega (\Delta R)^2}{\etaT},
\end{eqnarray}
where $\alpha$ is the turbulent $\alpha$ effect
\citep[e.g.][]{Steenbeck_et_al_1966_ZNatA_21_369,Brandenburg_Subramanian_2005_PhRp_417_1}
and $\etaT = \etat + \eta$ is the total magnetic diffusivity, where
$\etat$ is the turbulent contribution. An estimate of the
supercriticality in terms of the dynamo numbers can be obtained by
assuming that $\alpha$ and $\etat$ are independent of $\ReM$. This is
not the case for sufficiently low $\ReM$
\citep[e.g.][]{Sur_et_al_2008_MNRAS_385_L15}. Test field results from
Cartesian convection simulations suggest that for $\ReM\gtrsim 8$
\citep[][]{Kapyla_et_al_2009_AA_500_633} the turbulent dynamo
coefficients are indeed only weakly dependent of $\ReM$. However, no
corresponding test field calculations have been made with the current
set-up and therefore the values obtained here should be considered as
rough suggestive estimates. Using the dynamo number,
$D = C_{\alpha}C_{\Omega}$ as the relevant parameter with the
assumptions that $\alpha$ and $\etat$ are unaffected by $\ReM$, and
that $\etat/\eta \sim \ReM$, we find that in the kinematic regime
where the differential rotation is unaffected by magnetic fields, that
\begin{eqnarray}
\frac{D}{D_{\rm crit}} \equiv S_{\rm crit} = \left( \frac{1 + {\rm Re}_{\rm MC}^{-1}}{1+\ReM^{-1}} \right)^{2}.\label{equ:Scrit}
\end{eqnarray}
This means that $S_{\rm crit}$ varies between $0.90$ and $1.25$ in the
current runs; see the second row of \Table{tab:supercrit}.

Irrespective of the measure, the dynamos considered in the current
study are only mildly supercritical. Nevertheless, the change in
supercriticality in the parameter range considered here makes a
drastic difference in the behavior of the dynamo in the non-linear
regime. The current results support the idea that as stars age,
their dynamos become less supercritical and more prone to large
fluctuations and grand minima
\citep[e.g.][]{Baliunas_et_al_1995_ApJ_438_269,Schmitt_et_al_2026_AA_711_261}. However,
the frequency of grand minima in stars other than the Sun is still
under debate. Recent results of
\cite{Jarvinen_Strassmeier_2025_AA_698_93} suggest that such events
are very rare, contrary to the early conjecture of
\cite{Baliunas_et_al_1995_ApJ_438_269}.

\begin{table}[t!]
\centering
\caption[]{Supercriticality of the dynamo solutions.}
\label{tab:supercrit}
\vspace{-0.5cm}
$$
\begin{array}{rrrrrrrrrr}
  \hline
  \hline
  \noalign{\smallskip}
  \mbox{Run} & \mbox{A} & \mbox{B} & \mbox{C} & \mbox{D} & \mbox{E} & \mbox{F} & \mbox{G} & \mbox{H} & \mbox{I} \\
\hline
  R_{\rm crit} & 0.693 & 1.156 & 1.733 & 2.084 & 2.314 & 2.602 & 3.467 & 5.207 & 6.938 \\
  S_{\rm crit} & 0.899 & 1.034 & 1.113 & 1.142 & 1.156 & 1.171 & 1.201 & 1.233 & 1.250 \\
\hline
\end{array}
$$
\tablefoot{Supercriticality in terms of $\ReM$ ($R_{\rm crit}$) and
  $D$ ($S_{\rm crit}$) as defined in \Eqsa{equ:Rcrit}{equ:Scrit},
  respectively.}
\end{table}

\begin{figure*}
  \begin{center}
  \includegraphics[width=\textwidth]{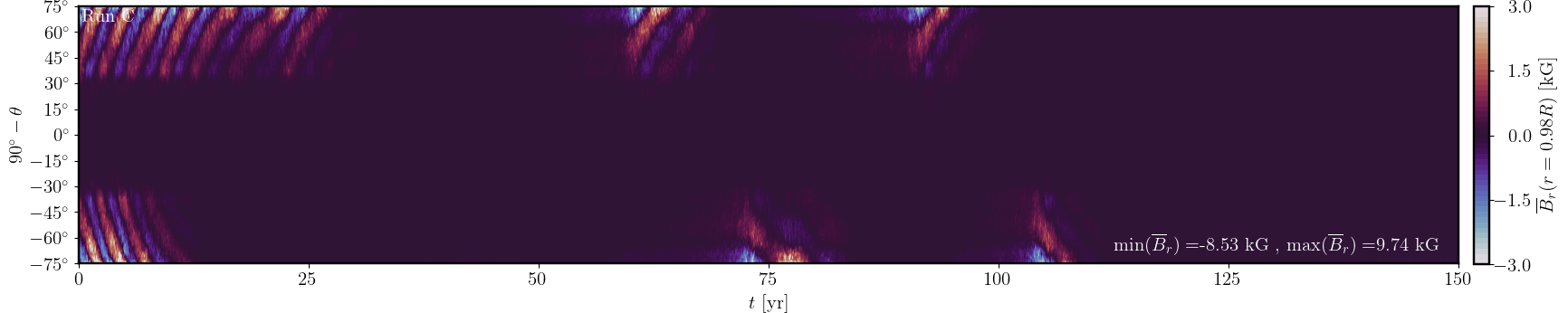}
  \includegraphics[width=\textwidth]{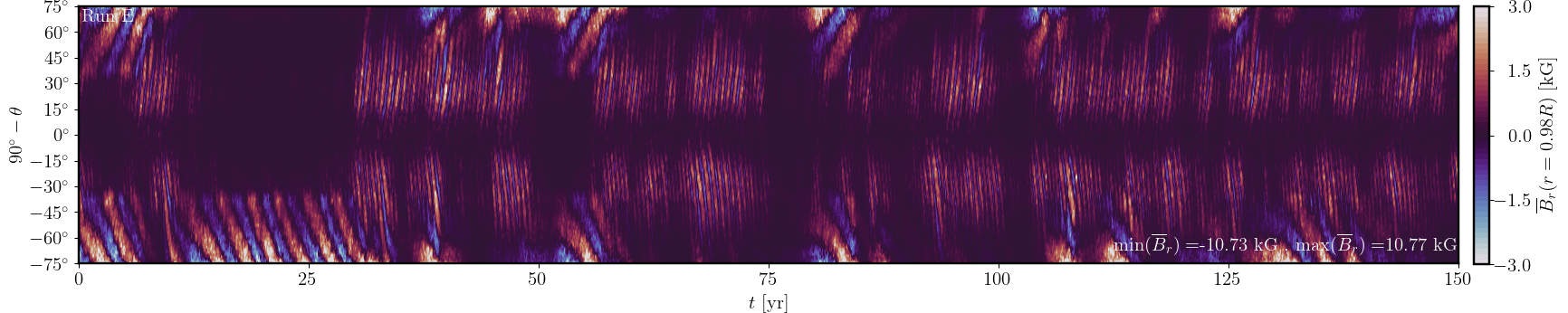}
  \includegraphics[width=\textwidth]{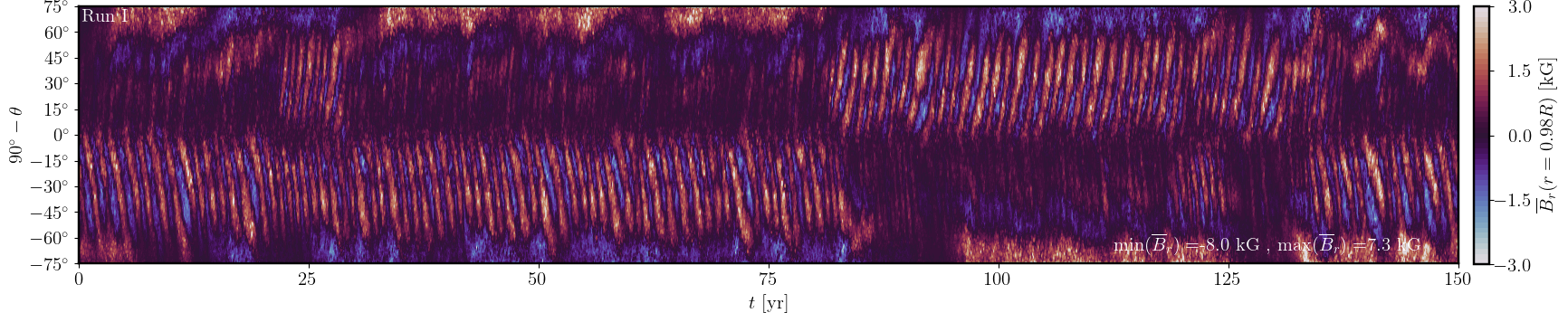}
  \end{center}
  \caption{Time-latitude diagrams of azimuthally averaged radial
    magnetic field $\mBBr(\theta,t)$ near the surface of the star at
    $r=0.98R$ for Runs~C (top), E (middle) and I (bottom) for a time
    period of 150~yr. The range of values is clipped in each panel and
    the minimum and maximum are indicated in the lower right corner of
    the figures.}
\label{fig:pbutter}
\end{figure*}

\subsection{Non-linear dynamo solutions}
\label{sec:dynsol}

Space-time diagrams of the azimuthally averaged radial magnetic field
$\mBBr(\theta,t)$ near the surface of the star at $r=0.98R$ for the
three representative runs are shown in \Figa{fig:pbutter}. In the
lowest $\PrM$ run (Run~C) an appreciable large-scale magnetic field is
only intermittently present and the strongest magnetic fields are
hemispheric at high latitudes. This dynamo mode shows poleward
migration at latitudes higher than about $\pm50\degr$. Practically no
magnetic fields are visible at latitudes below $\pm25\degr$.

In Run~E the high-latitude branch is still intermittently present but
another mode of intermittent activity appears at low latitudes
$0\degr \ldots \pm30\degr$. This mode is characterized by rapid cycles
and poleward migration. The field is on average less hemispherical
than in Run~C, although there are periods when one of the hemispheres
is clearly dominant; see, for example, the period between
10-30~yrs. This tendency is again more pronounced in Run~I where
hemispheric asymmetry is strong for the low-latitude poleward branch
which is now continuously present. In Run~I the high-latitude
oscillatory field of the lower-$\PrM$ runs is replaced by a more
quasi-static field that shows few polarity switches during the
simulated time period.

None of the simulations reproduce the solar butterfly diagram which is
a common issue with global stellar dynamo simulations
\citep[e.g.][]{Kapyla_et_al_2023_SSRv_219_58}. Earlier models with
set-ups similar to the current one in many respects
\cite[e.g.][]{Kapyla_et_al_2019_GAFD_113_149,Viviani_Kapyla_2021_AA_645_141}
achieved equatorward migrating activity near the equator. The main
difference is that a cooling layer instead of enhanced radiative
diffusion in the near-surface layers was used in those studies to
transport the luminosity through the upper boundary. This rather
subtle difference is enough to change the resulting flows such that
the most easily excited dynamo mode is quite different \citep[see
also][]{Kapyla_et_al_2020_GAFD_114_8}. Determining the precise cause
of the differences in the dynamo solutions requires extraction of
turbulent transport coefficients and detailed mean-field modeling
\citep[e.g.][]{Warnecke_et_al_2018_AA_609_51,Warnecke_et_al_2021_ApJL_919_13}
which is out of the scope of the current paper.

\subsection{Comparison to earlier simulations}

It is useful to put the current simulations in the context of earlier
solar and stellar dynamo simulations. Apparently most numerical
stellar dynamo studies occupy a regime where magnetic field is
constantly strong enough to quench the mean flows such that the
relative differential rotation typically ranges from a few to about
ten per cent \citep[e.g.][]{Kapyla_et_al_2012_ApJL_755_22,
  Kapyla_et_al_2013_ApJ_778_41, Guerrero_et_al_2019_ApJ_880_6,
  Matilsky_et_al_2026_ApJ_1005_134}. Such cases correspond to
Regime~III of the classification introduced earlier. Correspondingly
the variability of the large-scale magnetic fields is much weaker than
in the near-critical cases in Regime~I considered here
\citep[e.g.][]{Passos_Charbonneau_2014_AA_568_113,
  Augustson_et_al_2015_ApJ_809_149, Kapyla_MJ_et_al_2016_AA_589_56}.
Therefore most earlier dynamo simulations are likely in Regimes~II or
III explaining why grand minima-like events have so rare in earlier 3D
models.

Here the supercriticality of the dynamo is controlled by the imposed
magnetic diffusivity whereas in stars such as the Sun it is most
likely controlled by the strengths of the inductive effects such as
the $\alpha$ effect, resulting from helical flows, and differential
rotation. Both of these effects are sensitive to rotation such that
slower rotation typically leads to weaker supercriticality. This is
challenging to model with current spherical convection simulations
because they tend to generate stronger but anti-solar differential
rotation when the Coriolis number is less than about unity
\cite[e.g.][]{Gastine_et_al_2014_MNRAS_438_L76}, leading to enhanced
dynamo action \citep[][]{Karak_et_al_2015_AA_576_26,
  Warnecke_Kapyla_2020_AA_642_66}. Such behavior has not been verified
observationally, and it is possible that no anti-solar differential
rotation arises when sufficiently supercritical small-scale dynamo is
present \citep[][]{Hotta_Hatta_2026_NatAst_10_695}.

The effects of near-criticality arising from induction effects were
reported by \cite{Bushby_et_al_2018_AA_612_97} from Cartesian MHD
simulations of rotating convection. While rapidly rotating simulations
produced sustained large-scale dynamo action, intermittent bursts of
strong dynamo action as well as extended grand minima-like behavior
arose in weakly supercritical cases with slower rotation; see their
Figs.~10--12. In analogy to the current simulations, the strong-field
periods in the simulations of \cite{Bushby_et_al_2018_AA_612_97} were
characterized by the suppression of large-scale vortical flows that
were present in the weak-field and hydrodynamic regimes. Although the
study of \cite{Bushby_et_al_2018_AA_612_97} uses quite different model
assumptions, it is the closest parallel to the current study in that
weak supercriticality leads to more long-term variations reminiscent
of grand minima in the Sun.

\subsection{Relevance to the Sun and stars}
\label{sec:relevance}

The cycle-to-cycle variations of solar differential rotation are
predominantly seen in the form of torsional oscillations, the
magnitude of which is around one per cent \citep[e.g.][and references
therein]{Howe_2009_LRSP_6_1}. According to
\cite{Eddy_et_al_1976_SoPh_46_3} and
\cite{Eddy_et_al_1977_Science_198_824}, the solar equatorial rotation
was three to five per cent faster, and the latitudinal shear up to
three times larger during the Maunder minimum than
currently. \cite{Javaraiah_et_al_2005_SoPh_232_25} reported a roughly
linearly decreasing trend of solar differential rotation from cycles
12 to 23 which was associated with increasing solar activity. However,
these changes are also quite weak (of the order of one per cent) in
comparison to the variations seen in many of the current simulations;
see, for example, \Fig{fig:ptsuB3}. The relatively small variations in
solar differential rotation suggest that the solar dynamo is either in
a low or a high state, or in Regime~I or III, in the sense discussed
above. The intermediate, highly fluctuating Regime~II, is ruled out by
the observational evidence. The relative differential rotation in the
Sun is significantly stronger than in the current simulations. This
suggests that either the solar dynamo is in a low state in which the
differential rotation is almost unaffected by it (Regime~I), or the
current simulations fail to correctly capture the differential
rotation in the first place. Assuming that the former is the case, it
is possible to estimate from \Fig{fig:pcorruB} that the average
rms-magnetic field in the solar convection zone should be less than
about 2~kG for the differential rotation to remain strong. The
intermittency of the solar cycle over millennia
\citep[e.g.][]{Usoskin_2023_LRSP_20_2} would also be compatible with
the interpretation that the solar dynamo is in a low state
corresponding to Regime~I. Furthermore, \cite{Usoskin_2023_LRSP_20_2}
estimates that the current Sun spends 17 per cent of its time in a
grand minima. In the current Runs~B and C where grand minima
candidates are detected, their occurrence rate is roughly similar to
the solar data.

However, a major caveat in the current simulations is that the
magnetic Reynolds number is low, and a small-scale dynamo is absent
especially in the weakly supercritical cases. It is likely that an
efficient small-scale dynamo is present in the Sun
\citep{Warnecke_et_al_2023_NatAs_7_662}, and it is possible that it is
important in shaping the differential rotation
\citep{Hotta_Kusano_2021_NatAs_5_1100,Hotta_et_al_2022_ApJ_933_199,Hotta_2025_ApJ_985_163,Soderlund_et_al_2025_MNRAS_541_1816}. Therefore
a crucial test would be to run simulations where a supercritical
small-scale dynamo is present, but where the large-scale dynamo is
only marginally excited. If small-scale magnetic fields are crucial in
reaching solar-like differential rotation and large-scale dynamo at
solar $\CoF$, much higher resolution simulations than in the current
study need to be done to capture this
\citep[e.g.][]{Hotta_2025_ApJ_985_163}.

\section{Conclusions}
\label{sec:conclusions}

The current set of 3D simulations of convection-driven dynamos shows
that the overall level and fluctuations of both the flow and the
magnetic field are sensitive to the supercriticality of the dynamo
near the onset of dynamo action. Three distinct classes of solutions
were identified. Long periods, reaching several decades, of low
magnetic activity appear in cases where the dynamo is the least
supercritical (Regime~I). Such epochs are reminiscent of grand minima
of stellar magnetic activity, the most well-known being the Maunder
minimum of the Sun \citep[e.g.][]{Usoskin_2023_LRSP_20_2}; see also
\cite{Luhn_et_al_2022_ApJL_936_23} for a stellar example. In Regime~I
the flow is practically unaffected by the magnetic field, and shows
only weak fluctuations. Somewhat more supercritical runs (Regime~II)
exhibit strong fluctuations of both the velocity and magnetic field on
timescales ranging from a few years to about a decade. This occurs
because the magnetic field occasionally reaches a threshold value of
roughly 2~kG, above which it has a strong effect on the flow. The
resulting state of lower differential rotation is unable to maintain
the strong magnetic field and as the field decays the differential
rotation recovers. Finally, in the most supercritical cases
(Regime~III) the magnetic field is strong enough to suppress
differential rotation permanently and fluctuations of both the flow
and the magnetic field are weak.

The simulations of the current study do not reproduce solar
large-scale magnetism but the overall results can nevertheless have
implications for the Sun. First, the fact that weak supercriticality
leads to grand minimum-like behavior suggests that the solar dynamo is
only modestly supercritical. Mean-field modeling, albeit with quite
different assumptions about the dynamo, suggests a similar conclusion
\citep[][]{Wavhal_et_al_2025_SoPh_300_21}. Another independent hint
that the solar dynamo may be near the critical threshold and perhaps
fading away is the evidence of slowed magnetic braking of middle-aged
stars such as the Sun
\citep[][]{Metcalfe_et_al_2016_ApJL_826_2,van_Saders_et_al_2016_Nature_529_181}. Second,
the solar equatorial rotation rate varies typically less than a per
cent between cycle maximum and minimum \citep[e.g.][and references
therein]{Howe_2009_LRSP_6_1}, and estimates for the Maunder minimum
are of the order of three to four per cent
\citep[e.g.][]{Eddy_et_al_1977_Science_198_824}. The current
near-critical Regime~I simulations show differential rotation
variations of similar magnitude. Furthermore, the current results
indicate a strong anticorrelation between the rms-magnetic field and
the strength of differential rotation, and suggest that magnetic
fields below about 2~kG do not affect large-scale flows appreciably.

The current simulations were made in a regime where the small-scale
dynamo is likely absent. Small-scale magnetic fields from a highly
supercritical small-scale dynamo can have a major effect on the
large-scale flows
\citep[e.g.][]{Hotta_et_al_2022_ApJ_933_199,Soderlund_et_al_2025_MNRAS_541_1816},
and thus on the generation of magnetic fields. Future work needs to
explore parameter regimes where a supercritical small-scale dynamo is
present but where the large-scale dynamo remains nearly marginal to
confirm or to refute the astrophysical relevance of the current
results.

\begin{acknowledgements}
  PJK acknowledges financial support from the European Research
  Council through the ERC Advanced Grant NEOCON under grant
  No. 101199531. The simulations were made with resources granted by
  the Gauss Center for Supercomputing project ``Convection and dynamos
  near the transition to solar-like differential rotation (CONTRA)''
  in the Leibniz Supercomputing Centre's SuperMUC--NG supercomputer in
  Garching, Germany. OpenAI's Codex was used to improve the python
  scripts used to produce some of the figures and to track down
  inconsistencies, and typographic and grammar mistakes in the text.
\end{acknowledgements}

\bibliographystyle{aa}
\bibliography{master}

@ARTICLE{Augustson_et_al_2015_ApJ_809_149,
  author =	 {{Augustson}, K. and {Brun}, A.~S. and {Miesch},
                  M. and {Toomre}, J.},
  title =	 "{Grand Minima and Equatorward Propagation in a
                  Cycling Stellar Convective Dynamo}",
  journal =	 {\apj},
  archivePrefix ="arXiv",
  eprint =	 {1410.6547},
  primaryClass = "astro-ph.SR",
  year =	 2015,
  month =	 aug,
  volume =	 809,
  eid =		 {149},
  pages =	 {149},
  doi =		 {10.1088/0004-637X/809/2/149},
  adsurl =	 {http://adsabs.harvard.edu/abs/2015ApJ...809..149A},
}

@ARTICLE{Baliunas_et_al_1995_ApJ_438_269,
  author =	 {{Baliunas}, S.~L. and {Donahue}, R.~A. and {Soon},
                  W.~H. and {Horne}, J.~H. and {Frazer}, J. and
                  {Woodard-Eklund}, L. and {Bradford}, M. and {Rao},
                  L.~M. and {Wilson}, O.~C. and {Zhang}, Q. and
                  {Bennett}, W. and {Briggs}, J. and {Carroll},
                  S.~M. and {Duncan}, D.~K. and {Figueroa}, D. and
                  {Lanning}, H.~H. and {Misch}, T. and {Mueller},
                  J. and {Noyes}, R.~W. and {Poppe}, D. and {Porter},
                  A.~C. and {Robinson}, C.~R. and {Russell}, J. and
                  {Shelton}, J.~C. and {Soyumer}, T. and {Vaughan},
                  A.~H. and {Whitney}, J.~H.},
  title =	 "{Chromospheric Variations in Main-Sequence
                  Stars. II.}",
  journal =	 {\apj},
  year =	 "1995",
  month =	 "Jan",
  volume =	 {438},
  pages =	 {269},
  doi =		 {10.1086/175072},
  adsurl =
                  {https://ui.adsabs.harvard.edu/abs/1995ApJ...438..269B},
}

@ARTICLE{Bekki_2025_AA_703_262,
  author =	 {{Bekki}, Yuto},
  title =	 "{Impacts of small-scale dynamo on rotating columnar
                  convection in stellar convection zones}",
  journal =	 {\aap},
  year =	 2025,
  month =	 nov,
  volume =	 {703},
  eid =		 {A262},
  pages =	 {A262},
  doi =		 {10.1051/0004-6361/202556923},
  archivePrefix ={arXiv},
  eprint =	 {2509.19046},
  primaryClass = {astro-ph.SR},
  adsurl =
                  {https://ui.adsabs.harvard.edu/abs/2025A&A...703A.262B}
}

@ARTICLE{Brandenburg_Subramanian_2005_PhRp_417_1,
  author =	 {{Brandenburg}, A. and {Subramanian}, K.},
  title =	 "{Astrophysical magnetic fields and nonlinear dynamo
                  theory}",
  journal =	 {\physrep},
  eprint =	 {astro-ph/0405052},
  year =	 2005,
  month =	 oct,
  volume =	 417,
  pages =	 {1-209},
  doi =		 {10.1016/j.physrep.2005.06.005},
  adsurl =	 {http://esoads.eso.org/abs/2005PhR...417....1B},
}

@ARTICLE{Brandenburg_Spiegel_2008_AN_329_351,
  author =	 {{Brandenburg}, A. and {Spiegel}, E.~A.},
  title =	 "{Modeling a Maunder minimum}",
  journal =	 {Astronomische Nachrichten},
  year =	 2008,
  month =	 may,
  volume =	 {329},
  number =	 {4},
  pages =	 {351},
  doi =		 {10.1002/asna.200810973},
  archivePrefix ={arXiv},
  eprint =	 {0801.2156},
  primaryClass = {astro-ph},
  adsurl =
                  {https://ui.adsabs.harvard.edu/abs/2008AN....329..351B}
}

@ARTICLE{Pencil_Code_Collaboration_2021_JOSS_6_2807,
  author =	 {{Pencil Code Collaboration} and {Brandenburg}, Axel
                  and {Johansen}, Anders and {Bourdin}, Philippe and
                  {Dobler}, Wolfgang and {Lyra}, Wladimir and
                  {Rheinhardt}, Matthias and {Bingert}, Sven and
                  {Haugen}, Nils and {Mee}, Antony and {Gent},
                  Frederick and {Babkovskaia}, Natalia and {Yang},
                  Chao-Chin and {Heinemann}, Tobias and {Dintrans},
                  Boris and {Mitra}, Dhrubaditya and {Candelaresi},
                  Simon and {Warnecke}, J{\"o}rn and {K{\"a}pyl{\"a}},
                  Petri and {Schreiber}, Andreas and {Chatterjee},
                  Piyali and {K{\"a}pyl{\"a}}, Maarit and {Li},
                  Xiang-Yu and {Kr{\"u}ger}, Jonas and {Aarnes},
                  J{\o}rgen and {Sarson}, Graeme and {Oishi}, Jeffrey
                  and {Schober}, Jennifer and {Plasson}, Rapha{\"e}l
                  and {Sandin}, Christer and {Karchniwy}, Ewa and
                  {Rodrigues}, Luiz and {Hubbard}, Alexander and
                  {Guerrero}, Gustavo and {Snodin}, Andrew and
                  {Losada}, Illa and {Pekkil{\"a}}, Johannes and
                  {Qian}, Chengeng},
  title =	 "{The Pencil Code, a modular MPI code for partial
                  differential equations and particles: multipurpose
                  and multiuser-maintained}",
  journal =	 {The Journal of Open Source Software},
  year =	 2021,
  month =	 feb,
  volume =	 {6},
  number =	 {58},
  eid =		 {2807},
  pages =	 {2807},
  doi =		 {10.21105/joss.02807},
  adsurl =
                  {https://ui.adsabs.harvard.edu/abs/2021JOSS....6.2807P}
}

@ARTICLE{Brandenburg_et_al_2023_SSRv_219_55,
  author =	 {{Brandenburg}, Axel and {Elstner}, Detlef and
                  {Masada}, Youhei and {Pipin}, Valery},
  title =	 "{Turbulent Processes and Mean-Field Dynamo}",
  journal =	 {\ssr},
  year =	 2023,
  month =	 oct,
  volume =	 {219},
  number =	 {7},
  eid =		 {55},
  pages =	 {55},
  doi =		 {10.1007/s11214-023-00999-3},
  archivePrefix ={arXiv},
  eprint =	 {2303.12425},
  primaryClass = {astro-ph.SR},
  adsurl =
                  {https://ui.adsabs.harvard.edu/abs/2023SSRv..219...55B},
}

@ARTICLE{Brooke_et_al_2002_AA_395_1013,
  author =	 {{Brooke}, J. and {Moss}, D. and {Phillips}, A.},
  title =	 "{Deep minima in stellar dynamos}",
  journal =	 {\aap},
  year =	 2002,
  month =	 dec,
  volume =	 {395},
  pages =	 {1013-1022},
  doi =		 {10.1051/0004-6361:20021320},
  adsurl =
                  {https://ui.adsabs.harvard.edu/abs/2002A&A...395.1013B}
}

@ARTICLE{Bushby_2006_MNRAS_371_772,
  author =	 {{Bushby}, P.~J.},
  title =	 "{Zonal flows and grand minima in a solar dynamo
                  model}",
  journal =	 {\mnras},
  year =	 2006,
  month =	 sep,
  volume =	 {371},
  number =	 {2},
  pages =	 {772-780},
  doi =		 {10.1111/j.1365-2966.2006.10706.x},
  adsurl =
                  {https://ui.adsabs.harvard.edu/abs/2006MNRAS.371..772B}
}

@Article{Bushby_et_al_2018_AA_612_97,
  author =	 {{Bushby}, P.~J. and {K{\"a}pyl{\"a}}, P.~J. and
                  {Masada}, Y. and {Brandenburg}, A. and {Favier},
                  B. and {Guervilly}, C. and {K{\"a}pyl{\"a}}, M.~J.},
  journal =	 {\aap},
  title =	 {{Large-scale dynamos in rapidly rotating plane layer
                  convection}},
  year =	 {2018},
  month =	 may,
  pages =	 {A97},
  volume =	 {612},
  adsurl =	 {http://adsabs.harvard.edu/abs/2018A%26A...612A..97B},
  archiveprefix ={arXiv},
  doi =		 {10.1051/0004-6361/201732066},
  eid =		 {A97},
  eprint =	 {1710.03174},
  primaryclass = {astro-ph.SR},
}

@ARTICLE{Choudhuri_Karak_2012_PRL_109_171103,
  author =	 {{Choudhuri}, Arnab Rai and {Karak}, Bidya Binay},
  title =	 "{Origin of Grand Minima in Sunspot Cycles}",
  journal =	 {\prl},
  year =	 2012,
  month =	 oct,
  volume =	 {109},
  number =	 {17},
  eid =		 {171103},
  pages =	 {171103},
  doi =		 {10.1103/PhysRevLett.109.171103},
  archivePrefix ={arXiv},
  eprint =	 {1208.3947},
  primaryClass = {astro-ph.SR},
  adsurl =
                  {https://ui.adsabs.harvard.edu/abs/2012PhRvL.109q1103C}
}

@ARTICLE{Christensen_2002_JFM_470_115,
  author =	 {{Christensen}, U.~R.},
  title =	 "{Zonal flow driven by strongly supercritical
                  convection in rotating spherical shells}",
  journal =	 {Journal of Fluid Mechanics},
  year =	 2002,
  month =	 nov,
  volume =	 {470},
  number =	 {1},
  pages =	 {115-133},
  doi =		 {10.1017/S0022112002002008},
  adsurl =
                  {https://ui.adsabs.harvard.edu/abs/2002JFM...470..115C}
}

@ARTICLE{Christensen_Aubert_2006_GeJI_166_97,
  author =	 {{Christensen}, U.~R. and {Aubert}, J.},
  title =	 "{Scaling properties of convection-driven dynamos in
                  rotating spherical shells and application to
                  planetary magnetic fields}",
  journal =	 {Geophys. J. Int.},
  year =	 2006,
  month =	 jul,
  volume =	 166,
  pages =	 {97-114},
  doi =		 {10.1111/j.1365-246X.2006.03009.x},
  adsurl =	 {http://esoads.eso.org/abs/2006GeoJI.166...97C}
}

@ARTICLE{Dobler_et_al_2006_ApJ_638_336,
  author =	 {{Dobler}, W. and {Stix}, M. and {Brandenburg}, A.},
  title =	 "{Magnetic Field Generation in Fully Convective
                  Rotating Spheres}",
  journal =	 {\apj},
  eprint =	 {arXiv:astro-ph/0410645},
  year =	 2006,
  month =	 feb,
  volume =	 638,
  pages =	 {336-347},
  doi =		 {10.1086/498634},
  adsurl =	 {http://adsabs.harvard.edu/abs/2006ApJ...638..336D},
}

@article{Eddy_1976_Science_192_4245,
  author =	 {Eddy, John A.},
  title =	 {The Maunder Minimum},
  volume =	 {192},
  number =	 {4245},
  pages =	 {1189--1202},
  year =	 {1976},
  doi =		 {10.1126/science.192.4245.1189},
  publisher =	 {American Association for the Advancement of Science},
  issn =	 {0036-8075},
  URL =
                  {https://science.sciencemag.org/content/192/4245/1189},
  eprint =
                  {https://science.sciencemag.org/content/192/4245/1189.full.pdf},
  journal =	 {Science}
}

@ARTICLE{Eddy_et_al_1976_SoPh_46_3,
  author =	 {{Eddy}, J.~A. and {Gilman}, P.~A. and {Trotter},
                  D.~E.},
  title =	 "{Solar rotation during the Maunder minimum.}",
  journal =	 {\solphys},
  year =	 1976,
  month =	 jan,
  volume =	 {46},
  number =	 {1},
  pages =	 {3-14},
  doi =		 {10.1007/BF00157550},
  adsurl =
                  {https://ui.adsabs.harvard.edu/abs/1976SoPh...46....3E}
}

@ARTICLE{Eddy_et_al_1977_Science_198_824,
  author =	 {{Eddy}, J.~A. and {Gilman}, P.~A. and {Trotter},
                  D.~E.},
  title =	 "{Anomalous Solar Rotation in the Early 17th
                  Century}",
  journal =	 {Science},
  year =	 1977,
  month =	 nov,
  volume =	 {198},
  number =	 {4319},
  pages =	 {824-829},
  doi =		 {10.1126/science.198.4319.824},
  adsurl =
                  {https://ui.adsabs.harvard.edu/abs/1977Sci...198..824E}
}

@ARTICLE{Gastine_et_al_2014_MNRAS_438_L76,
  author =	 {{Gastine}, T. and {Yadav}, R.~K. and {Morin}, J. and
                  {Reiners}, A. and {Wicht}, J.},
  title =	 "{From solar-like to antisolar differential rotation
                  in cool stars}",
  journal =	 {\mnras},
  archivePrefix ="arXiv",
  eprint =	 {1311.3047},
  primaryClass = "astro-ph.SR",
  year =	 2014,
  month =	 feb,
  volume =	 438,
  pages =	 {L76-L80},
  doi =		 {10.1093/mnrasl/slt162},
  adsurl =	 {http://adsabs.harvard.edu/abs/2014MNRAS.438L..76G},
}

@ARTICLE{Ghizaru_et_al_2010_ApJL_715_133,
  author =	 {{Ghizaru}, M. and {Charbonneau}, P. and
                  {Smolarkiewicz}, P.~K.},
  title =	 "{Magnetic Cycles in Global Large-eddy Simulations of
                  Solar Convection}",
  journal =	 {\apjl},
  year =	 2010,
  month =	 jun,
  volume =	 715,
  pages =	 {L133-L137},
  doi =		 {10.1088/2041-8205/715/2/L133},
  adsurl =	 {http://adsabs.harvard.edu/abs/2010ApJ...715L.133G}
}

@ARTICLE{Guerrero_et_al_2019_ApJ_880_6,
  author =	 {{Guerrero}, G. and {Zaire}, B. and {Smolarkiewicz},
                  P.~K. and {de Gouveia Dal Pino}, E.~M. and
                  {Kosovichev}, A.~G. and {Mansour}, N.~N.},
  title =	 "{What Sets the Magnetic Field Strength and Cycle
                  Period in Solar-type Stars?}",
  journal =	 {\apj},
  year =	 "2019",
  month =	 "Jul",
  volume =	 {880},
  number =	 {1},
  eid =		 {6},
  pages =	 {6},
  doi =		 {10.3847/1538-4357/ab224a},
  archivePrefix ={arXiv},
  eprint =	 {1810.07978},
  primaryClass = {astro-ph.SR},
  adsurl =
                  {https://ui.adsabs.harvard.edu/abs/2019ApJ...880....6G}
}

@ARTICLE{Hathaway_2015_LRSP_12_4,
  author =	 {{Hathaway}, David H.},
  title =	 "{The Solar Cycle}",
  journal =	 {Living Reviews in Solar Physics},
  year =	 2015,
  month =	 sep,
  volume =	 {12},
  number =	 {1},
  eid =		 {4},
  pages =	 {4},
  doi =		 {10.1007/lrsp-2015-4},
  archivePrefix ={arXiv},
  eprint =	 {1502.07020},
  primaryClass = {astro-ph.SR},
  adsurl =
                  {https://ui.adsabs.harvard.edu/abs/2015LRSP...12....4H}
}

@ARTICLE{Hotta_Kusano_2021_NatAs_5_1100,
  author =	 {{Hotta}, H. and {Kusano}, K.},
  title =	 "{Solar differential rotation reproduced with
                  high-resolution simulation}",
  journal =	 {Nature Astronomy},
  year =	 2021,
  month =	 sep,
  volume =	 {5},
  pages =	 {1100-1102},
  doi =		 {10.1038/s41550-021-01459-0},
  archivePrefix ={arXiv},
  eprint =	 {2109.06280},
  primaryClass = {astro-ph.SR},
  adsurl =
                  {https://ui.adsabs.harvard.edu/abs/2021NatAs...5.1100H},
}

@ARTICLE{Hotta_et_al_2022_ApJ_933_199,
  author =	 {{Hotta}, H. and {Kusano}, K. and {Shimada}, R.},
  title =	 "{Generation of Solar-like Differential Rotation}",
  journal =	 {\apj},
  year =	 2022,
  month =	 jul,
  volume =	 {933},
  number =	 {2},
  eid =		 {199},
  pages =	 {199},
  doi =		 {10.3847/1538-4357/ac7395},
  archivePrefix ={arXiv},
  eprint =	 {2202.04183},
  primaryClass = {astro-ph.SR},
  adsurl =
                  {https://ui.adsabs.harvard.edu/abs/2022ApJ...933..199H},
}

@ARTICLE{Hotta_2025_ApJ_985_163,
  author =	 {{Hotta}, H.},
  title =	 "{Simultaneous Construction of Fast Equator, Poleward
                  Meridional Flow, and Near-surface Shear Layer in
                  Solar Magnetohydrodynamic Calculation}",
  journal =	 {\apj},
  year =	 2025,
  month =	 jun,
  volume =	 {985},
  number =	 {2},
  eid =		 {163},
  pages =	 {163},
  doi =		 {10.3847/1538-4357/adca3b},
  archivePrefix ={arXiv},
  eprint =	 {2504.05680},
  primaryClass = {astro-ph.SR},
  adsurl =
                  {https://ui.adsabs.harvard.edu/abs/2025ApJ...985..163H}
}

@ARTICLE{Hotta_Hatta_2026_NatAst_10_695,
  author =	 {{Hotta}, Hideyuki and {Hatta}, Yoshiki},
  title =	 "{The prevalence of solar-like differential rotation
                  in slowly rotating solar-type stars}",
  journal =	 {Nature Astronomy},
  year =	 2026,
  month =	 may,
  volume =	 {10},
  pages =	 {695-701},
  doi =		 {10.1038/s41550-026-02793-x},
  adsurl =
                  {https://ui.adsabs.harvard.edu/abs/2026NatAs..10..695H}
}

@ARTICLE{Howe_2009_LRSP_6_1,
  author =	 {{Howe}, Rachel},
  title =	 "{Solar Interior Rotation and its Variation}",
  journal =	 {Living Reviews in Solar Physics},
  year =	 2009,
  month =	 dec,
  volume =	 {6},
  number =	 {1},
  eid =		 {1},
  pages =	 {1},
  doi =		 {10.12942/lrsp-2009-1},
  archivePrefix ={arXiv},
  eprint =	 {0902.2406},
  primaryClass = {astro-ph.SR},
  adsurl =
                  {https://ui.adsabs.harvard.edu/abs/2009LRSP....6....1H},
}

@ARTICLE{Javaraiah_et_al_2005_SoPh_232_25,
  author =	 {{Javaraiah}, J. and {Bertello}, L. and {Ulrich},
                  R.~K.},
  title =	 "{Long-Term Variations in Solar Differential Rotation
                  and Sunspot Activity}",
  journal =	 {\solphys},
  year =	 2005,
  month =	 nov,
  volume =	 {232},
  number =	 {1-2},
  pages =	 {25-40},
  doi =		 {10.1007/s11207-005-8776-y},
  adsurl =
                  {https://ui.adsabs.harvard.edu/abs/2005SoPh..232...25J}
}

@ARTICLE{Jarvinen_Strassmeier_2025_AA_698_93,
  author =	 {{J{\"a}rvinen}, S.~P. and {Strassmeier}, K.~G.},
  title =	 "{A search for Maunder-minimum candidate stars}",
  journal =	 {\aap},
  year =	 2025,
  month =	 jun,
  volume =	 {698},
  eid =		 {A93},
  pages =	 {A93},
  doi =		 {10.1051/0004-6361/202554111},
  archivePrefix ={arXiv},
  eprint =	 {2504.19670},
  primaryClass = {astro-ph.SR},
  adsurl =
                  {https://ui.adsabs.harvard.edu/abs/2025A&A...698A..93J}
}

@ARTICLE{Karak_et_al_2015_AA_576_26,
  author =	 {{Karak}, B.~B. and {K{\"a}pyl{\"a}}, P.~J. and
                  {K{\"a}pyl{\"a}}, M.~J. and {Brandenburg}, A. and
                  {Olspert}, N. and {Pelt}, J.},
  title =	 "{Magnetically controlled stellar differential
                  rotation near the transition from solar to
                  anti-solar profiles}",
  journal =	 {\aap},
  archivePrefix ="arXiv",
  eprint =	 {1407.0984},
  primaryClass = "astro-ph.SR",
  year =	 2015,
  month =	 apr,
  volume =	 576,
  eid =		 {A26},
  pages =	 {A26},
  doi =		 {10.1051/0004-6361/201424521},
  adsurl =	 {http://esoads.eso.org/abs/2015A%26A...576A..26K}
}

@ARTICLE{Karak_2023_LRSP_20_3,
  author =	 {{Karak}, Bidya Binay},
  title =	 "{Models for the long-term variations of solar
                  activity}",
  journal =	 {Living Reviews in Solar Physics},
  year =	 2023,
  month =	 dec,
  volume =	 {20},
  number =	 {1},
  eid =		 {3},
  pages =	 {3},
  doi =		 {10.1007/s41116-023-00037-y},
  archivePrefix ={arXiv},
  eprint =	 {2305.17188},
  primaryClass = {astro-ph.SR},
  adsurl =
                  {https://ui.adsabs.harvard.edu/abs/2023LRSP...20....3K}
}

@ARTICLE{Kupka_Muthsam_2017_LRCA_3_1,
  author =	 {{Kupka}, F. and {Muthsam}, H.~J.},
  title =	 "{Modelling of stellar convection}",
  journal =	 {Liv. Rev. Comp. Astrophys.},
  year =	 2017,
  month =	 jul,
  volume =	 3,
  eid =		 {1},
  pages =	 {1},
  doi =		 {10.1007/s41115-017-0001-9},
  adsurl =	 {http://adsabs.harvard.edu/abs/2017LRCA....3....1K},
}

@ARTICLE{Kapyla_MJ_et_al_2016_AA_589_56,
  author =	 {{K{\"a}pyl{\"a}}, M.~J. and {K{\"a}pyl{\"a}},
                  P.~J. and {Olspert}, N. and {Brandenburg}, A. and
                  {Warnecke}, J. and {Karak}, B.~B. and {Pelt}, J.},
  title =	 "{Multiple dynamo modes as a mechanism for long-term
                  solar activity variations}",
  journal =	 {\aap},
  archivePrefix ="arXiv",
  eprint =	 {1507.05417},
  primaryClass = "astro-ph.SR",
  year =	 2016,
  month =	 apr,
  volume =	 589,
  eid =		 {A56},
  pages =	 {A56},
  doi =		 {10.1051/0004-6361/201527002},
  adsurl =	 {http://adsabs.harvard.edu/abs/2016A%26A...589A..56K}
}

@Article{Kapyla_et_al_2009_AA_500_633,
  author =	 {{K{\"a}pyl{\"a}}, P.~J. and {Korpi}, M.~J. and
                  {Brandenburg}, A.},
  journal =	 {\aap},
  title =	 {{Alpha effect and turbulent diffusion from
                  convection}},
  year =	 {2009},
  month =	 jun,
  pages =	 {633-646},
  volume =	 {500},
  adsurl =	 {http://adsabs.harvard.edu/abs/2009A%26A...500..633K},
  archiveprefix ={arXiv},
  doi =		 {10.1051/0004-6361/200811498},
  eprint =	 {0812.1792},
}

@Article{Kapyla_et_al_2012_ApJL_755_22,
  author =	 {{K{\"a}pyl{\"a}}, P.~J. and {Mantere}, M.~J. and
                  {Brandenburg}, A.},
  journal =	 {\apjl},
  title =	 {{Cyclic Magnetic Activity due to Turbulent
                  Convection in Spherical Wedge Geometry}},
  year =	 {2012},
  month =	 aug,
  pages =	 {L22},
  volume =	 {755},
  adsurl =	 {http://adsabs.harvard.edu/abs/2012ApJ...755L..22K},
  archiveprefix ={arXiv},
  doi =		 {10.1088/2041-8205/755/1/L22},
  eid =		 {L22},
  eprint =	 {1205.4719},
  primaryclass = {astro-ph.SR},
}

@Article{Kapyla_et_al_2013_ApJ_778_41,
  author =	 {{K{\"a}pyl{\"a}}, P.~J. and {Mantere}, M.~J. and
                  {Cole}, E. and {Warnecke}, J. and {Brandenburg}, A.},
  journal =	 {\apj},
  title =	 {{Effects of Enhanced Stratification on Equatorward
                  Dynamo Wave Propagation}},
  year =	 {2013},
  month =	 nov,
  pages =	 {41},
  volume =	 {778},
  adsurl =	 {http://adsabs.harvard.edu/abs/2013ApJ...778...41K},
  archiveprefix ={arXiv},
  doi =		 {10.1088/0004-637X/778/1/41},
  eid =		 {41},
  eprint =	 {1301.2595},
  primaryclass = {astro-ph.SR},
}

@Article{Kapyla_et_al_2014_AA_570_43,
  author =	 {{K{\"a}pyl{\"a}}, P.~J. and {K{\"a}pyl{\"a}},
                  M.~J. and {Brandenburg}, A.},
  journal =	 {\aap},
  title =	 {{Confirmation of bistable stellar differential
                  rotation profiles}},
  year =	 {2014},
  month =	 oct,
  pages =	 {A43},
  volume =	 {570},
  adsurl =	 {http://adsabs.harvard.edu/abs/2014arXiv1401.2981K},
  archiveprefix ={arXiv},
  eid =		 {A43},
  eprint =	 {1401.2981},
  primaryclass = {astro-ph.SR},
}

@ARTICLE{Kapyla_et_al_2017_AA_599_4,
  author =	 {{K{\"a}pyl{\"a}}, P.~J. and {K{\"a}pyl{\"a}},
                  M.~J. and {Olspert}, N. and {Warnecke}, J. and
                  {Brandenburg}, A.},
  title =	 "{Convection-driven spherical shell dynamos at
                  varying Prandtl numbers}",
  journal =	 {\aap},
  archivePrefix ="arXiv",
  eprint =	 {1605.05885},
  primaryClass = "astro-ph.SR",
  year =	 2017,
  month =	 mar,
  volume =	 599,
  eid =		 {A4},
  pages =	 {A4},
  doi =		 {10.1051/0004-6361/201628973},
  adsurl =
                  {http://cdsads.u-strasbg.fr/abs/2017A%26A...599A...4K},
}

@ARTICLE{Kapyla_2019_AA_622_195,
  author =	 {{K{\"a}pyl{\"a}}, P.~J.},
  title =	 "{Magnetic and rotational quenching of the $\Lambda$
                  effect}",
  journal =	 {\aap},
  archivePrefix ="arXiv",
  eprint =	 {1712.08045},
  primaryClass = "astro-ph.SR",
  year =	 2019,
  volume =	 622,
  pages =	 {A195},
  month =	 dec,
  adsurl =	 {http://cdsads.u-strasbg.fr/abs/2017arXiv171208045K},
}

@ARTICLE{Kapyla_et_al_2019_GAFD_113_149,
  author =	 {{K{\"a}pyl{\"a}}, P.~J. and {Viviani}, M. and
                  {K{\"a}pyl{\"a}}, M.~J. and {Brandenburg}, A. and
                  {Spada}, F.},
  title =	 "{Effects of a subadiabatic layer on convection and
                  dynamos in spherical wedge simulations}",
  journal =	 {Geophysical and Astrophysical Fluid Dynamics},
  archivePrefix ="arXiv",
  eprint =	 {1803.05898},
  primaryClass = "astro-ph.SR",
  year =	 2019,
  month =	 mar,
  volume =	 113,
  pages =	 {149-183},
  doi =		 {10.1080/03091929.2019.1571584},
  adsurl =
                  {https://ui.adsabs.harvard.edu/abs/2019GApFD.113..149K},
}

@ARTICLE{Kapyla_et_al_2020_GAFD_114_8,
  author =	 {{K{\"a}pyl{\"a}}, P.~J. and {Gent}, F.~A. and
                  {Olspert}, N. and {K{\"a}pyl{\"a}}, M.~J. and
                  {Brandenburg}, A.},
  title =	 "{Sensitivity to luminosity, centrifugal force, and
                  boundary conditions in spherical shell convection}",
  journal =	 {Geophysical and Astrophysical Fluid Dynamics},
  year =	 "2020",
  month =	 "Mar",
  volume =	 {114},
  number =	 {1-2},
  pages =	 {8-34},
  doi =		 {10.1080/03091929.2019.1571586},
  adsurl =
                  {https://ui.adsabs.harvard.edu/abs/2020GApFD.114....8K},
}

@ARTICLE{Kapyla_2022_ApJ_931_L17,
  author =	 {{K{\"a}pyl{\"a}}, Petri J.},
  title =	 "{Solar-like dynamos and rotational scaling of cycles
                  from star-in-a-box simulations}",
  journal =	 {\apjl},
  volume =	 931,
  eid =		 {L17},
  pages =	 {L17},
  doi =		 {10.3847/2041-8213/ac6e6b},
  year =	 2022,
  month =	 jun,
  eid =		 {arXiv:2202.04329},
  pages =	 {arXiv:2202.04329},
  archivePrefix ={arXiv},
  eprint =	 {2202.04329},
  primaryClass = {astro-ph.SR}
}

@ARTICLE{Kapyla_2023_AA_669_98,
  author =	 {{K{\"a}pyl{\"a}}, P.~J.},
  title =	 "{Transition from anti-solar to solar-like
                  differential rotation: Dependence on Prandtl
                  number}",
  journal =	 {\aap},
  year =	 2023,
  month =	 jan,
  volume =	 {669},
  eid =		 {A98},
  pages =	 {A98},
  doi =		 {10.1051/0004-6361/202244395},
  archivePrefix ={arXiv},
  eprint =	 {2207.00302},
  primaryClass = {astro-ph.SR},
  adsurl =
                  {https://ui.adsabs.harvard.edu/abs/2023A&A...669A..98K}
}

@ARTICLE{Kapyla_et_al_2023_SSRv_219_58,
  author =	 {{K{\"a}pyl{\"a}}, Petri J. and {Browning}, Matthew
                  K. and {Brun}, Allan Sacha and {Guerrero}, Gustavo
                  and {Warnecke}, J{\"o}rn},
  title =	 "{Simulations of Solar and Stellar Dynamos and Their
                  Theoretical Interpretation}",
  journal =	 {\ssr},
  year =	 2023,
  month =	 oct,
  volume =	 {219},
  number =	 {7},
  eid =		 {58},
  pages =	 {58},
  doi =		 {10.1007/s11214-023-01005-6},
  archivePrefix ={arXiv},
  eprint =	 {2305.16790},
  primaryClass = {astro-ph.SR},
  adsurl =
                  {https://ui.adsabs.harvard.edu/abs/2023SSRv..219...58K}
}

@ARTICLE{Kapyla_2024_AA_683_221,
  author =	 {{K{\"a}pyl{\"a}}, P.~J.},
  title =	 "{Convective scale and subadiabatic layers in
                  simulations of rotating compressible convection}",
  journal =	 {\aap},
  year =	 2024,
  month =	 mar,
  volume =	 {683},
  eid =		 {A221},
  pages =	 {A221},
  doi =		 {10.1051/0004-6361/202348325},
  archivePrefix ={arXiv},
  eprint =	 {2310.12855},
  primaryClass = {astro-ph.SR},
  adsurl =
                  {https://ui.adsabs.harvard.edu/abs/2024A&A...683A.221K}
}

@ARTICLE{Kapyla_2025_LRSP_22_3,
  author =	 {{K{\"a}pyl{\"a}}, Petri J.},
  title =	 "{Connecting mean-field theory with dynamo
                  simulations}",
  journal =	 {Living Reviews in Solar Physics},
  year =	 2025,
  month =	 sep,
  volume =	 {22},
  number =	 {1},
  eid =		 {3},
  pages =	 {3},
  doi =		 {10.1007/s41116-025-00042-3},
  archivePrefix ={arXiv},
  eprint =	 {2507.00632},
  primaryClass = {astro-ph.SR},
  adsurl =
                  {https://ui.adsabs.harvard.edu/abs/2025LRSP...22....3K}
}

@ARTICLE{Luhn_et_al_2022_ApJL_936_23,
  author =	 {{Luhn}, Jacob K. and {Wright}, Jason T. and {Henry},
                  Gregory W. and {Saar}, Steven H. and {Baum}, Anna
                  C.},
  title =	 "{HD 166620: Portrait of a Star Entering a Grand
                  Magnetic Minimum}",
  journal =	 {\apjl},
  year =	 2022,
  month =	 sep,
  volume =	 {936},
  number =	 {2},
  eid =		 {L23},
  pages =	 {L23},
  doi =		 {10.3847/2041-8213/ac8b13},
  archivePrefix ={arXiv},
  eprint =	 {2207.00612},
  primaryClass = {astro-ph.SR},
  adsurl =
                  {https://ui.adsabs.harvard.edu/abs/2022ApJ...936L..23L}
}

@ARTICLE{Matilsky_et_al_2026_ApJ_1005_134,
  author =	 {{Matilsky}, Loren I. and {Korre}, Lydia and
                  {Brummell}, Nicholas H.},
  title =	 "{A Dynamo Confinement Scenario for the Solar
                  Tachocline and Its Implications for Spin-down in the
                  Radiative Spreading Regime}",
  journal =	 {\apj},
  year =	 2026,
  month =	 jul,
  volume =	 {1005},
  number =	 {2},
  eid =		 {134},
  pages =	 {134},
  doi =		 {10.3847/1538-4357/ae6705},
  archivePrefix ={arXiv},
  eprint =	 {2601.11943},
  primaryClass = {astro-ph.SR},
  adsurl =
                  {https://ui.adsabs.harvard.edu/abs/2026ApJ..1005..134M}
}

@ARTICLE{Maunder_1894_KIMS_18_173,
  author =	 {{Maunder}, E. Walter},
  title =	 "{A Prolonged Sunspot Minimum}",
  journal =	 {Knowledge: An Illustrated Magazine of Science},
  year =	 "1894",
  month =	 "Aug",
  volume =	 {17},
  pages =	 {173-176},
  adsurl =
                  {https://ui.adsabs.harvard.edu/abs/1894KIMS...18..173M},
}

@ARTICLE{Metcalfe_et_al_2016_ApJL_826_2,
  author =	 {{Metcalfe}, T.~S. and {Egeland}, R. and {van
                  Saders}, J.},
  title =	 "{Stellar Evidence That the Solar Dynamo May Be in
                  Transition}",
  journal =	 {\apjl},
  archivePrefix ="arXiv",
  eprint =	 {1606.01926},
  primaryClass = "astro-ph.SR",
  year =	 2016,
  month =	 jul,
  volume =	 826,
  eid =		 {L2},
  pages =	 {L2},
  doi =		 {10.3847/2041-8205/826/1/L2},
  adsurl =	 {http://adsabs.harvard.edu/abs/2016ApJ...826L...2M}
}

@Article{Mitra_et_al_2009_ApJ_697_923,
  author =	 {{Mitra}, D. and {Tavakol}, R. and {Brandenburg},
                  A. and {Moss}, D.},
  journal =	 {\apj},
  title =	 {{Turbulent Dynamos in Spherical Shell Segments of
                  Varying Geometrical Extent}},
  year =	 {2009},
  month =	 may,
  pages =	 {923-933},
  volume =	 {697},
  adsurl =	 {http://adsabs.harvard.edu/abs/2009ApJ...697..923M},
  archiveprefix ={arXiv},
  doi =		 {10.1088/0004-637X/697/1/923},
  eprint =	 {0812.3106},
}

@ARTICLE{Oelcek_et_al_2019_SoPh_294_99,
  author =	 {{{\"O}l{\c{c}}ek}, Deniz and {Charbonneau}, Paul and
                  {Lemerle}, Alexandre and {Longpr{\'e}}, Gabriel and
                  {Boileau}, Florence},
  title =	 "{Grand Activity Minima and Maxima via Dual Dynamos}",
  journal =	 {\solphys},
  year =	 2019,
  month =	 jul,
  volume =	 {294},
  number =	 {7},
  eid =		 {99},
  pages =	 {99},
  doi =		 {10.1007/s11207-019-1492-9},
  adsurl =
                  {https://ui.adsabs.harvard.edu/abs/2019SoPh..294...99O}
}

@ARTICLE{OMara_et_al_2016_AdSpR_58_1475,
  author =	 {{O'Mara}, B. and {Miesch}, M.~S. and {Featherstone},
                  N.~A. and {Augustson}, K.~C.},
  title =	 "{Velocity amplitudes in global convection
                  simulations: The role of the Prandtl number and
                  near-surface driving}",
  journal =	 {Adv. Space Res.},
  archivePrefix ="arXiv",
  eprint =	 {1603.06107},
  primaryClass = "astro-ph.SR",
  year =	 2016,
  month =	 oct,
  volume =	 58,
  pages =	 {1475-1489},
  doi =		 {10.1016/j.asr.2016.03.038},
  adsurl =	 {http://adsabs.harvard.edu/abs/2016AdSpR..58.1475O},
}

@ARTICLE{Ossendrijver_2000_AA_359_364,
  author =	 {{Ossendrijver}, M.~A.~J.~H.},
  title =	 "{Grand minima in a buoyancy-driven solar dynamo}",
  journal =	 {\aap},
  year =	 2000,
  month =	 jul,
  volume =	 {359},
  pages =	 {364-372},
  adsurl =
                  {https://ui.adsabs.harvard.edu/abs/2000A&A...359..364O}
}

@ARTICLE{Passos_Charbonneau_2014_AA_568_113,
  author =	 {{Passos}, D. and {Charbonneau}, P.},
  title =	 "{Characteristics of magnetic solar-like cycles in a
                  3D MHD simulation of solar convection}",
  journal =	 {\aap},
  year =	 2014,
  month =	 aug,
  volume =	 568,
  eid =		 {A113},
  pages =	 {A113},
  doi =		 {10.1051/0004-6361/201423700},
  adsurl =	 {http://adsabs.harvard.edu/abs/2014A%26A...568A.113P}
}

@ARTICLE{Schmitt_et_al_2026_AA_711_261,
  author =	 {{Schmitt}, J.~H.~M.~M. and {Schr{\"o}der}, K.-P. and
                  {Mittag}, M. and {Hempelmann}, A. and
                  {Gonz{\'a}lez-P{\'e}rez}, J.~N. and {Jack}, D.},
  title =	 "{Six decades of TIGRE and Mount Wilson chromospheric
                  monitoring in the H and K lines: The quest for an
                  understanding of solar-type activity}",
  journal =	 {\aap},
  year =	 2026,
  month =	 jul,
  volume =	 {711},
  eid =		 {A261},
  pages =	 {A261},
  doi =		 {10.1051/0004-6361/202557956},
  archivePrefix ={arXiv},
  eprint =	 {2606.26283},
  primaryClass = {astro-ph.SR},
  adsurl =
                  {https://ui.adsabs.harvard.edu/abs/2026A&A...711A.261S}
}

@ARTICLE{Schumacher_Sreenivasan_2020_RvMP_92_041001,
  author =	 {{Schumacher}, J{\"o}rg and {Sreenivasan}, Katepalli
                  R.},
  title =	 "{Colloquium: Unusual dynamics of convection in the
                  Sun}",
  journal =	 {Reviews of Modern Physics},
  year =	 2020,
  month =	 oct,
  volume =	 {92},
  number =	 {4},
  eid =		 {041001},
  pages =	 {041001},
  doi =		 {10.1103/RevModPhys.92.041001},
  adsurl =
                  {https://ui.adsabs.harvard.edu/abs/2020RvMP...92d1001S}
}

@ARTICLE{Soderlund_et_al_2025_MNRAS_541_1816,
  author =	 {{Soderlund}, Krista M. and {Wulff}, Paula and
                  {K{\"a}pyl{\"a}}, Petri J. and {Aurnou}, Jonathan
                  M.},
  title =	 "{Magnetohydrodynamic control of differential
                  rotation and dynamo transitions: rise of the local
                  magnetic Rossby number}",
  journal =	 {\mnras},
  year =	 2025,
  month =	 aug,
  volume =	 {541},
  number =	 {2},
  pages =	 {1816-1830},
  doi =		 {10.1093/mnras/staf1081},
  adsurl =
                  {https://ui.adsabs.harvard.edu/abs/2025MNRAS.541.1816S}
}

@ARTICLE{Solanki_et_al_2004_Nature_431_1084,
  author =	 {{Solanki}, S.~K. and {Usoskin}, I.~G. and {Kromer},
                  B. and {Sch{\"u}ssler}, M. and {Beer}, J.},
  title =	 "{Unusual activity of the Sun during recent decades
                  compared to the previous 11,000 years}",
  journal =	 {\nat},
  year =	 "2004",
  month =	 "Oct",
  volume =	 {431},
  number =	 {7012},
  pages =	 {1084-1087},
  doi =		 {10.1038/nature02995},
  adsurl =
                  {https://ui.adsabs.harvard.edu/abs/2004Natur.431.1084S}
}

@ARTICLE{Steenbeck_et_al_1966_ZNatA_21_369,
  author =	 {{Steenbeck}, M. and {Krause}, F. and {R{\"a}dler},
                  K.-H.},
  title =	 "{Berechnung der mittleren Lorentz-Feldst{\"a}rke
                  $\overline{\bm v \times \bm{B}}$ f{\"u}r ein
                  elektrisch leitendes Medium in turbulenter, durch
                  Coriolis-Kr{\"a}fte beeinflu{\ss}ter Bewegung}",
  journal =	 {Zeitschrift Naturforschung Teil A},
  year =	 1966,
  month =	 apr,
  volume =	 21,
  pages =	 {369},
  doi =		 {10.1515/zna-1966-0401},
  adsurl =	 {http://adsabs.harvard.edu/abs/1966ZNatA..21..369S}
}

@Article{Strugarek_et_al_2018_ApJ_863_35,
  author =	 {{Strugarek}, A. and {Beaudoin}, P. and
                  {Charbonneau}, P. and {Brun}, A.~S.},
  journal =	 {\apj},
  title =	 {{On the Sensitivity of Magnetic Cycles in Global
                  Simulations of Solar-like Stars}},
  year =	 {2018},
  month =	 aug,
  pages =	 {35},
  volume =	 {863},
  adsurl =	 {http://adsabs.harvard.edu/abs/2018ApJ...863...35S},
  archiveprefix ={arXiv},
  doi =		 {10.3847/1538-4357/aacf9e},
  eid =		 {35},
  eprint =	 {1806.09484},
  primaryclass = {astro-ph.SR}
}

@ARTICLE{Sur_et_al_2008_MNRAS_385_L15,
  author =	 {{Sur}, S. and {Brandenburg}, A. and {Subramanian},
                  K.},
  title =	 "{Kinematic {$\alpha$}-effect in isotropic turbulence
                  simulations}",
  journal =	 {\mnras},
  archivePrefix ="arXiv",
  eprint =	 {0711.3789},
  year =	 2008,
  month =	 mar,
  volume =	 385,
  pages =	 {L15-L19},
  doi =		 {10.1111/j.1745-3933.2008.00423.x},
  adsurl =	 {http://esoads.eso.org/abs/2008MNRAS.385L..15S}
}

@ARTICLE{Tobias_1996_AA_307_L21,
  author =	 {{Tobias}, S.~M.},
  title =	 "{Grand minimia in nonlinear dynamos.}",
  journal =	 {\aap},
  year =	 1996,
  month =	 mar,
  volume =	 {307},
  pages =	 {L21},
  adsurl =
                  {https://ui.adsabs.harvard.edu/abs/1996A&A...307L..21T},
}

@ARTICLE{Usoskin_2023_LRSP_20_2,
  author =	 {{Usoskin}, Ilya G.},
  title =	 "{A history of solar activity over millennia}",
  journal =	 {Living Reviews in Solar Physics},
  year =	 2023,
  month =	 dec,
  volume =	 {20},
  number =	 {1},
  eid =		 {2},
  pages =	 {2},
  doi =		 {10.1007/s41116-023-00036-z},
  adsurl =
                  {https://ui.adsabs.harvard.edu/abs/2023LRSP...20....2U}
}

@ARTICLE{van_Saders_et_al_2016_Nature_529_181,
  author =	 {{van Saders}, J.~L. and {Ceillier}, T. and
                  {Metcalfe}, T.~S. and {Silva Aguirre}, V. and
                  {Pinsonneault}, M.~H. and {Garc{\'{\i}}a}, R.~A. and
                  {Mathur}, S. and {Davies}, G.~R.},
  title =	 "{Weakened magnetic braking as the origin of
                  anomalously rapid rotation in old field stars}",
  journal =	 {\nat},
  archivePrefix ="arXiv",
  eprint =	 {1601.02631},
  primaryClass = "astro-ph.SR",
  year =	 2016,
  month =	 jan,
  volume =	 529,
  pages =	 {181-184},
  doi =		 {10.1038/nature16168},
  adsurl =	 {http://adsabs.harvard.edu/abs/2016Natur.529..181V}
}

@ARTICLE{Vashishth_et_al_2023_MNRAS_522_2601,
  author =	 {{Vashishth}, Vindya and {Karak}, Bidya Binay and
                  {Kitchatinov}, Leonid},
  title =	 "{Dynamo modelling for cycle variability and
                  occurrence of grand minima in Sun-like stars:
                  rotation rate dependence}",
  journal =	 {\mnras},
  year =	 2023,
  month =	 jun,
  volume =	 {522},
  number =	 {2},
  pages =	 {2601-2610},
  doi =		 {10.1093/mnras/stad1105},
  archivePrefix ={arXiv},
  eprint =	 {2304.05819},
  primaryClass = {astro-ph.SR},
  adsurl =
                  {https://ui.adsabs.harvard.edu/abs/2023MNRAS.522.2601V}
}

@ARTICLE{Viviani_Kapyla_2021_AA_645_141,
  author =	 {{Viviani}, M. and {K{\"a}pyl{\"a}}, M.~J.},
  title =	 "{Physically motivated heat-conduction treatment in
                  simulations of solar-like stars: effects on dynamo
                  transitions}",
  journal =	 {\aap},
  year =	 2021,
  month =	 jan,
  volume =	 {645},
  eid =		 {A141},
  pages =	 {A141},
  doi =		 {10.1051/0004-6361/202038603},
  archivePrefix ={arXiv},
  eprint =	 {2006.04426},
  primaryClass = {astro-ph.SR},
  adsurl =
                  {https://ui.adsabs.harvard.edu/abs/2021A&A...645A.141V}
}

@ARTICLE{Warnecke_2018_AA_616_72,
  author =	 {{Warnecke}, J.},
  title =	 "{Dynamo cycles in global convection simulations of
                  solar-like stars}",
  journal =	 {\aap},
  archivePrefix ="arXiv",
  eprint =	 {1712.01248},
  primaryClass = "astro-ph.SR",
  year =	 2018,
  month =	 aug,
  volume =	 616,
  eid =		 {A72},
  pages =	 {A72},
  doi =		 {10.1051/0004-6361/201732413}
}

@ARTICLE{Warnecke_et_al_2018_AA_609_51,
  author =	 {{Warnecke}, J. and {Rheinhardt}, M. and {Tuomisto},
                  S. and {K{\"a}pyl{\"a}}, P.~J. and {K{\"a}pyl{\"a}},
                  M.~J. and {Brandenburg}, A.},
  title =	 "{Turbulent transport coefficients in spherical wedge
                  dynamo simulations of solar-like stars}",
  journal =	 {\aap},
  archivePrefix ="arXiv",
  eprint =	 {1601.03730},
  primaryClass = "astro-ph.SR",
  year =	 2018,
  month =	 jan,
  volume =	 609,
  eid =		 {A51},
  pages =	 {A51},
  doi =		 {10.1051/0004-6361/201628136}
}

@ARTICLE{Warnecke_Kapyla_2020_AA_642_66,
  author =	 {{Warnecke}, J. and {K{\"a}pyl{\"a}}, M.~J.},
  title =	 "{Rotational dependence of turbulent transport
                  coefficients in global convective dynamo simulations
                  of solar-like stars}",
  journal =	 {\aap},
  year =	 2020,
  month =	 oct,
  volume =	 {642},
  eid =		 {A66},
  pages =	 {A66},
  doi =		 {10.1051/0004-6361/201936922},
  archivePrefix ={arXiv},
  eprint =	 {1910.06776},
  primaryClass = {astro-ph.SR},
  adsurl =
                  {https://ui.adsabs.harvard.edu/abs/2020A&A...642A..66W}
}

@ARTICLE{Warnecke_et_al_2021_ApJL_919_13,
  author =	 {{Warnecke}, J{\"o}rn and {Rheinhardt}, Matthias and
                  {Viviani}, Mariangela and {Gent}, Frederick A. and
                  {Tuomisto}, Simo and {K{\"a}pyl{\"a}}, Maarit J.},
  title =	 "{Investigating Global Convective Dynamos with
                  Mean-field Models: Full Spectrum of Turbulent
                  Effects Required}",
  journal =	 {\apjl},
  year =	 2021,
  month =	 oct,
  volume =	 {919},
  number =	 {2},
  eid =		 {L13},
  pages =	 {L13},
  doi =		 {10.3847/2041-8213/ac1db5},
  archivePrefix ={arXiv},
  eprint =	 {2105.07708},
  primaryClass = {astro-ph.SR},
  adsurl =
                  {https://ui.adsabs.harvard.edu/abs/2021ApJ...919L..13W}
}

@ARTICLE{Warnecke_et_al_2023_NatAs_7_662,
  author =	 {{Warnecke}, J{\"o}rn and {Korpi-Lagg}, Maarit J. and
                  {Gent}, Frederick A. and {Rheinhardt}, Matthias},
  title =	 "{Numerical evidence for a small-scale dynamo
                  approaching solar magnetic Prandtl numbers}",
  journal =	 {Nature Astronomy},
  year =	 2023,
  month =	 jun,
  volume =	 {7},
  pages =	 {662-668},
  doi =		 {10.1038/s41550-023-01975-1},
  archivePrefix ={arXiv},
  eprint =	 {2306.03991},
  primaryClass = {astro-ph.SR},
  adsurl =
                  {https://ui.adsabs.harvard.edu/abs/2023NatAs...7..662W}
}

@ARTICLE{Warnecke_et_al_2025_AA_696_93,
  author =	 {{Warnecke}, J. and {Korpi-Lagg}, M.~J. and
                  {Rheinhardt}, M. and {Viviani}, M. and {Prabhu}, A.},
  title =	 "{Small-scale and large-scale dynamos in global
                  convection simulations of solar-like stars}",
  journal =	 {\aap},
  year =	 2025,
  month =	 apr,
  volume =	 {696},
  eid =		 {A93},
  pages =	 {A93},
  doi =		 {10.1051/0004-6361/202451085},
  archivePrefix ={arXiv},
  eprint =	 {2406.08967},
  primaryClass = {astro-ph.SR},
  adsurl =
                  {https://ui.adsabs.harvard.edu/abs/2025A&A...696A..93W}
}

@ARTICLE{Wavhal_et_al_2025_SoPh_300_21,
  author =	 {{Wavhal}, Sanket and {Kumar}, Pawan and {Karak},
                  Bidya Binay},
  title =	 "{Analyses of Features of Magnetic Cycles at
                  Different Amounts of Dynamo Supercriticality: Solar
                  Dynamo Is About Two Times Critical}",
  journal =	 {\solphys},
  year =	 2025,
  month =	 feb,
  volume =	 {300},
  number =	 {2},
  eid =		 {21},
  pages =	 {21},
  doi =		 {10.1007/s11207-025-02428-w},
  archivePrefix ={arXiv},
  eprint =	 {2501.02262},
  primaryClass = {astro-ph.SR},
  adsurl =
                  {https://ui.adsabs.harvard.edu/abs/2025SoPh..300...21W}
}

@Book{Weiss_et_al_2004_Cox_Giuli_Principles,
  author =	 {{Weiss}, A. and {Hillebrandt}, W. and {Thomas},
                  H.-C. and {Ritter}, H.},
  publisher =	 {Cambridge Scientific Publishers Ltd},
  title =	 {{Cox and Giuli's Principles of Stellar Structure}},
  year =	 {2004},
  address =	 {Cambridge, UK},
  adsurl =	 {http://adsabs.harvard.edu/abs/2004cgps.book.....W},
  booktitle =	 {Cox and Giuli's Principles of Stellar Structure, by
                  A.~Weiss, W.~Hillebrandt, H-C.~Thomas,
                  H.~Ritter.~Cambridge, UK: Princeton Publishing
                  Associates Ltd, 2004.}
}

\end{document}